\def\puncspace{\ifmmode\,\else{\ifcat.\C{\if.\C\else\if,\C\else\if?\C\else%
\if:\C\else\if;\C\else\if-\C\else\if)\C\else\if/\C\else\if]\C\else\if'\C%
\else\space\fi\fi\fi\fi\fi\fi\fi\fi\fi\fi}%
\else\if\empty\C\else\if\space\C\else\space\fi\fi\fi}\fi}
\def\SP{\let\\=\empty\futurelet\C\puncspace}

\def\h1{$h^{-1}$\SP}

\def\etal{{\it et al.\/}\ }

\def\eg{{\it e.g.\/}\rm,\ }
\def\lsim{~\rlap{$<$}{\lower 1.0ex\hbox{$\sim$}}}
\def\gsim{~\rlap{$>$}{\lower 1.0ex\hbox{$\sim$}}}
\def\void#1{{}}
\def\u{$U$}
\def\b{$B$}
\def\v{$V$}
\def\r{$R$}
\def\i{$I$}
\def\j{$J$}
\def\h{$H$}
\def\k{$Ks$}
\documentclass{aa}
\usepackage{graphicx}

\usepackage{times}
\usepackage{mathptm}
\fontfamily{ptmcm}\selectfont
\usepackage{graphicx}

\begin{document}

   \thesaurus{02 (04.19.1; 04.03.1; 08.07.1; 11.07.1; 12.03.3)}
   \title{ESO Imaging Survey}

   \subtitle{Hubble Deep Field South: Optical-Infrared Observations,
Data Reduction and Photometry}

\author{  
       L. da Costa\inst{1}
 \and M. Nonino \inst{1,2}
 \and R. Rengelink \inst{1}
 \and S. Zaggia \inst{1,3,4} 
 \and C. Benoist\inst{1,5} 
 \and T. Erben\inst{1,5}
 \and A. Wicenec\inst{1}
 \and M. Scodeggio \inst{1}
 \and L. F. Olsen\inst{1,6}
 \and D. Guarnieri\inst{7} 
 \and E. Deul\inst{1,8}
 \and S. D'Odorico \inst{1}
 \and R. Hook\inst{9} 
 \and A. Moorwood \inst{1}
 \and R. Slijkhuis\inst{1}  
 }

\institute{
European Southern Observatory, Karl-Schwarzschild-Str. 2,
D--85748 Garching b. M\"unchen, Germany 
 \and Osservatorio Astronomico di Trieste, Via G.B. Tiepolo 11, I-31144
Trieste, Italy
\and Dipartimento di Astronomia, Univ. di Padova, vicolo
dell'Osservatorio 5, I-35125, Padova, Italy
\and Osservatorio Astronomico di Capodimonte, via Moiariello 15,
 I-80131, Napoli, Italy 
 \and Max-Planck Institut f\"ur Astrophysik, Postfach 1523 D-85748, 
 Garching bei M\"unchen, Germany
 \and Astronomisk Observatorium, Juliane Maries Vej 30, DK-2100 Copenhagen, 
  Denmark
 \and Osservatorio Astronomico di Pino Torinese, Strada Osservatorio
 20, I-10025 Torino, Italy
  \and Leiden Observatory, P.O. Box 9513, 2300 RA Leiden, The
  Netherlands 
  \and Space Telescope -- European Coordinating Facility,
 Karl-Schwarzschild-Str. 2, D--85748 Garching b. M\"unchen, Germany }




   \date{Received ; accepted }

   \maketitle
    

   \begin{abstract}

This paper presents ground-based data obtained from deep optical and
infrared observations of the HST Hubble Deep Field South (HDF-S) field
carried out at the ESO 3.5 New Technology Telescope (NTT).  These data
were taken as part of the ESO Imaging Survey (EIS) program, a public
survey coordinated by ESO and member states, in preparation for the
first year of operation of the VLT. Deep CCD images are available for
five optical passbands, reaching $2\sigma$ limiting magnitudes of
$U_{AB}\sim$ 27.0, $B_{AB}\sim$ 26.5, $V_{AB}\sim$ 26, $R_{AB}\sim$26,
$I_{AB}\sim$ 25, covering a region of $\sim$ 25 square arcmin, which
includes the HST WPFC2 field.  The infrared observations cover a total
area of $\sim$ 42 square arcmin and include both the HST WFPC2 and
STIS fields. The observations of the WFPC2 region were conducted in
$JHKs$ passbands, reaching $J_{AB}\sim 25$, and $H_{AB}$ and $K_{AB}\sim$
24.0. Due to time constraints, the adjacent field, covering the STIS
field, has been observed only in $R$, $I$ and $JHKs$, while no
observations were conducted covering the NIC3 field. This paper
describes the observations and data reduction. It also presents images
of the surveyed region and lists the optical and infrared photometric
parameters of the objects detected on the co-added images of each
passband, as well as multicolor optical and infrared catalogs.  These
catalogs together with the astrometrically and photometrically
calibrated co-added images are being made public world-wide and can be
retrieved from the world-wide web (http://www.eso.org/eis).

   \end{abstract}

%

\section{Introduction}

One of the main goals of the ESO Imaging Survey (EIS, Renzini \& da
Costa 1997) has been to carry out deep, multicolor observations in the
optical and infrared passbands over a relatively large area ($\sim$
200 square arcmin) to produce faint galaxy samples (EIS-DEEP). The
primary objective is to use the color information to estimate
photometric redshifts, and identify galaxies likely to be in the
poorly sampled $1 \lsim z<2$ redshift interval or Lyman-break
candidates at $z\gsim2.5$, all interesting targets for follow-up
spectroscopic observations with the VLT.

Following the remarkable success of the HST observations of the Hubble
Deep Field (HDF) in the North, the STScI has now completed a similar
campaign in a second field, this time accessible from
southern-hemisphere facilities. For HDF-S, the HST carried out
simultaneous observations with three of its major instruments (WFPC2,
STIS, and NICMOS) in parallel observing modes, leading to deep images
of three separate fields. Another important difference relative to the
original HDF is that instead of using an "undistinguished" field, the
selected pointing for HDF-S was chosen to contain a QSO within the
STIS field. Although this choice is interesting in many respects, the
selected field is close to a very bright star as well as several to
relatively bright stars, making it less than ideal for wide-angle,
deep ground-based optical and near-infrared observations.

To reconcile the great interest generated by the HDF-S with the desire
to conduct observations in a less crowded field, the EIS Working Group
recommended the EIS-DEEP observations to be split into two sets: one
set consisting of three adjacent fields, of about 25 square arcmin
each, covering the WFPC2, STIS and NIC3 fields; and another set
consisting of four adjacent fields, comprising a total area of 100
square arcmin. covering a region of very low  HI column density at
$\alpha=03^h32^m28^s $ and $\delta=-27^\circ48'30'$. This field is
particularly interesting as deep imaging in X-ray of the region is
planned with AXAF.

\begin{figure*}
\resizebox{\textwidth}{!}{\includegraphics{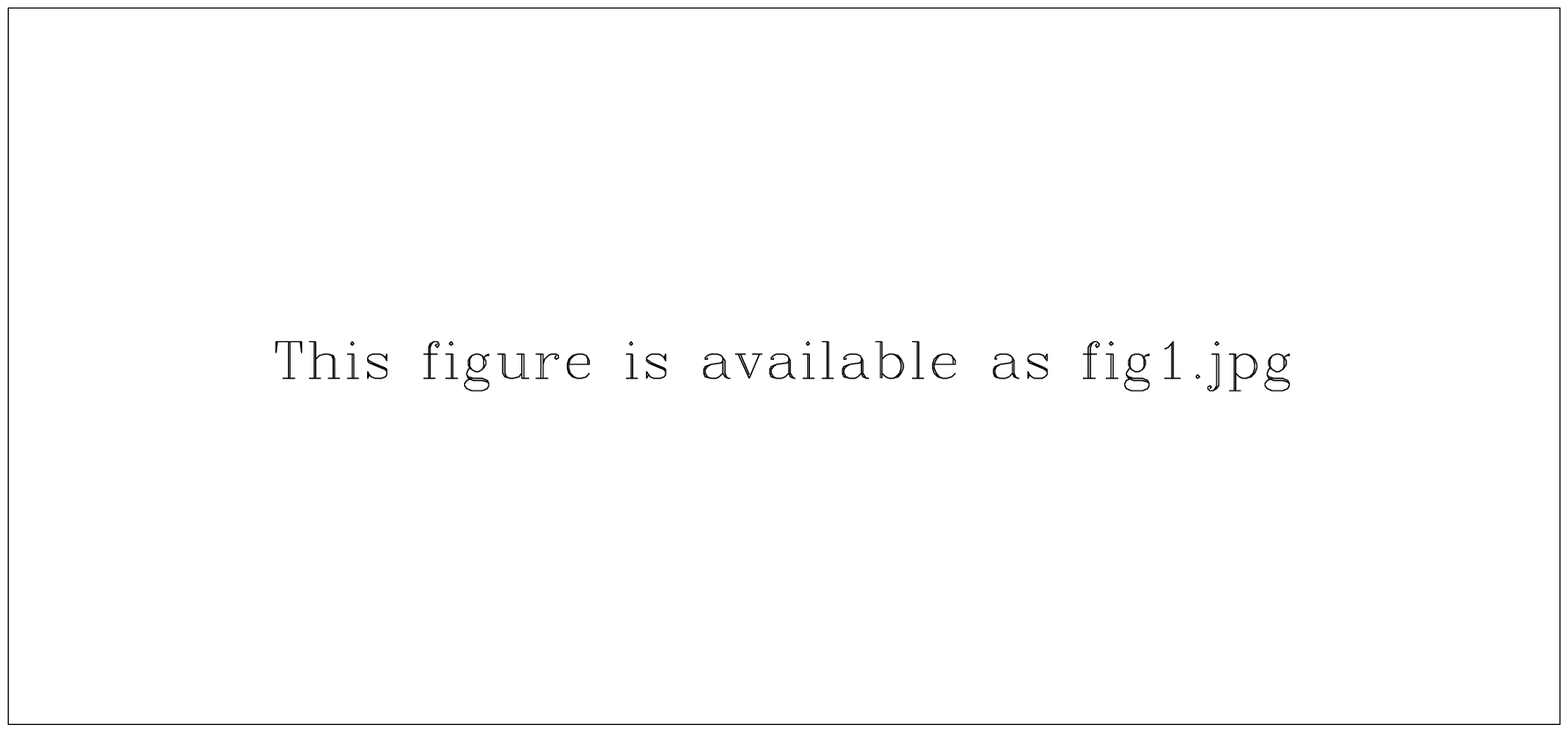}}
\caption{EIS-DEEP observations of the HDF-S. The image corresponds to
the co-added \r-band image, the only optical image currently available
covering both the HST-WFPC2 and STIS fields, depicted in the figure.
Also shown are the positions of the two SUSI2 chips (the paired
rectangles) and the SOFI coverage (the large squares) at the nominal
pointings listed in Table~\ref{tab:pointings}. The SUSI2 gap and the
SOFI overlap can be clearly seen. For all images shown in this
paper north is up and east to the left.}
\label{fig:hdfs}
\end{figure*}

In the present paper the data from observations of the HDF-S field,
conducted in five optical and three infrared passbands over the period
August-November 1998, are reported.  In section~\ref{obs}, the
observations and data reduction are discussed, and co-added images for
the different fields and passbands are presented. In
section~\ref{cats}, the basic photometric parameters of sources
detected in each passband are listed. In addition. optical, infrared
and optical-infrared multicolor catalogs, based on detections in a
combined reference image are presented. In section~\ref{comp}, a
comparison of the data is made with other observations of the same
field to evaluate the completeness and the accuracy of the magnitude
measurements. Using both the single-band and multi-color catalogs the
characteristics of the data are explored in section~\ref{results}
where galaxy and star number counts as well as color-color diagrams
are presented. Finally, a preliminary list of high-z galaxy candidates
is selected.  A brief summary is presented in section~\ref{sum}.

\section {Observations and Data Reduction}
\label{obs}

\subsection{Observing Strategy}

The original goal of the EIS-DEEP observations of the HDF-S fields was
to observe three adjacent, and preferably contiguous, regions
overlapping the WFPC2, STIS and the NIC3 fields both in optical and
infrared. Therefore, the pointings of the optical and infrared cameras
were chosen based on the relative separation of these fields and the
physical characteristics of the cameras, thereby constraining the
final sky coverage and the overlap between the optical and infrared
observations.

\begin{table}
\caption{SUSI2 and SOFI Pointings}
\begin {tabular}{lcccc}
\hline \hline
Field & $\alpha_{susi2}$ & $\delta_{susi2}$ & $\alpha_{sofi}$ &
$\delta_{sofi}$ \\
\hline
HDF1 & 22:33:29.1 &$-60$:33:50 & 22:33:32.5 &$-60$:33:30 \\
HDF2 & 22:32:42.4 &$-60$:33:50 & 22:33:00.0 &$-60$:33:30 \\
\hline
\hline
\label{tab:pointings}
\end{tabular}
\end{table}

The optical observations were carried out using the SUSI2 camera
(D'Odorico \etal 1998) at the f/11 Nasmyth focus A of the New
Technology Telescope (NTT). The camera consists of two thinned,
anti-reflection coated, $2k\times4k$, $15\mu m$ pixel EEV CCDs (ESO
\#45, and \#46), with the long side aligned in the north-south
direction, leading to a field of view of $5.46\times5.46$ square
arcmin.  The pixel scale of the camera is $0.08$ arcsec/pixel but the
observations were carried out using a $2\times2$ binning, yielding a
scale of $0.16$ arcsec per pixel. On the sky the gap separating the
two CCDs corresponds to $\sim 8$ arcsec.  The observations were
carried out in a dithered pattern, designed to minimize the lack of
sensitivity at the center of the camera due to the gap.  The dithering
pattern adopted for SUSI2 consisted of eleven pointings, evenly
distributed within a rectangular box 40 arcsec long in right ascension
and 8~arcsec wide in declination. The region where the sensitivity of
the combined images is $\gsim 70\%$ is considered below to be the
effective size of the observations in a given passband.  After the
combination of the 11 pointings, this area includes the gap and a
small region close to the outer edge of the frame, where the
sensitivity drops to about 70\% or $\sim$0.25 mag brighter than the
limit reached in regions covered by all frames. The effective area
covered by one SUSI2 observation is therefore $5.20\times5.40$ square
arcmin, corresponding to 94\% of the SUSI2 field of view in a single
exposure.  Using the weight map computed by the EIS pipeline, a mask
for each passband was created which accurately reflects the effective
area.  However, since there may be small relative shifts of the
centers of the co-added images in different passbands, a master mask
has also been created (the product of the individual band masks), to
identify all the pixels that are above a minimum sensitivity ($\gsim$
70\%) in all bands. Light streaks, associated with stars outside the
field of view, have also been masked out. The final effective area
used to trim all the co-added images is about $5.3 \times 5.3$ square
arcmin.  All catalogs presented below have been compiled using objects
extracted within this common area, except in the \u-band, where an
additional region was masked out to avoid the effects associated with
the wings of a bright star in the field.

\begin{table}
\caption{Summary of Optical Observations}
\begin{tabular}
{lrcccc}
\hline\hline
Filter & $t_{total}$ &  $N_f$ & seeing       & FWHM  & $\mu_{lim}$\\
       &             &        &        range &       &            \\
       &    (sec)    &        &  (arcsec)    & (arcsec) &  (mag arcsec$^{-2}$) \\
\hline
HDF1       &      &       &           &       \\
\hline
  \r &  5500 &   22  & 0.8-1.3 & 1.15 & 28.01\\
  \i &  3600 &    6  & 1.3-1.5 & 1.50 & 26.86 \\
\hline
HDF2       &      &       &           &       & \\
\hline
   \u & 17800 &   22  & 0.7-1.2 & 1.00    & 27.87 \\
   \b &  6600 &   22  & 0.7-1.2 & 0.84    & 28.88\\
   \v & 12250 &   49  & 0.9-1.5 & 1.27    & 28.89\\
   \r &  5500 &   22  & 0.8-1.3 & 1.05    & 28.01\\
   \i &  8800 &   44  & 0.9-1.4 & 1.11    & 27.23\\
\hline\hline
\end{tabular}
\label{tab:exposures}
\end{table}

\begin{figure}
\resizebox{0.45\textwidth}{!}{\includegraphics{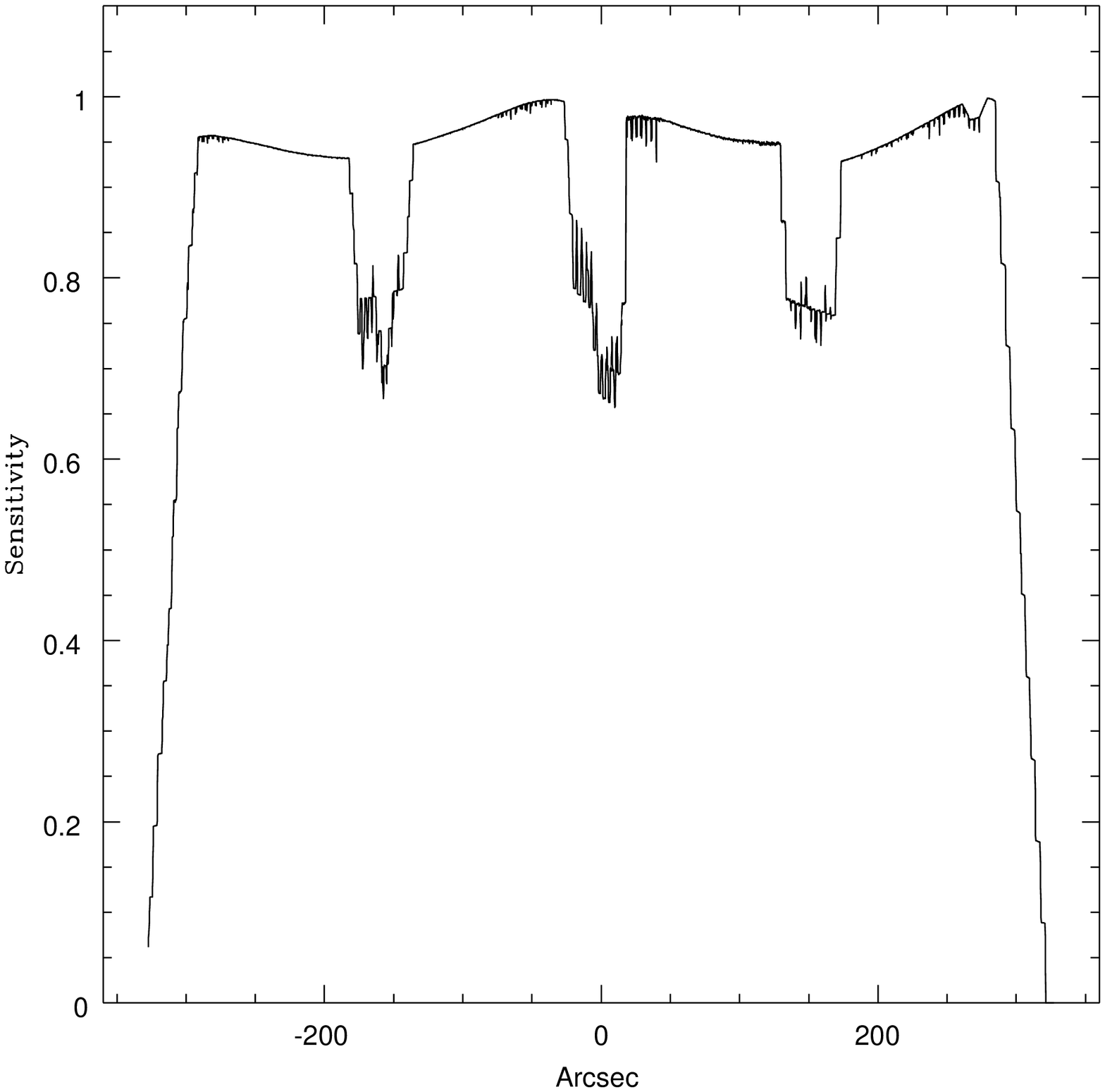}}
\caption{Cut across the normalized weight map of the fully  co-added
SUSI2 \r-band image shown in Figure~\ref{fig:hdfs}. The figure shows
the variation of the sensitivity across the image along the right
ascension direction. Note how the dithering fills the gap between the
chips and between the pointings.}
\label{fig:sensitivity}
\end{figure}

Since the size of the HST-WFPC2 field of view ($\sim158$ arcsec) is
comparable to the short axis of one SUSI2 chip ($\sim164$ arcsec), the
nominal center of the SUSI2 observations of the WFPC2 field was chosen
so that the latter would lie within a single SUSI2 chip, with the
center slightly shifted to the west to avoid as much as possible the
lack of sensitivity due to the gap. The original idea was that the
lack of sensitivity at the eastern edge of the chip, also affected by
the dithering, would be compensated by exposures of the adjacent field
covering the STIS field. Taking these constraints into consideration
the reference pointings finally adopted for SUSI2 are listed in
Table~\ref{tab:pointings} for the two fields observed.  To illustrate
the geometry of the EIS observations of HDF-S, Figure~\ref{fig:hdfs}
shows the co-added $R-$band image built from the images taken in the
direction of the STIS and WFPC2 fields, hereafter referred to as HDF1
and HDF2.  Also shown are the locations of the two SUSI2 chips when
the centers listed in Table~\ref{tab:pointings} are used.  For
comparison the approximate locations of the HST-WFPC2 and STIS fields
are also displayed. Complementing this information,
Figure~\ref{fig:sensitivity} shows a cut at fixed declination of the
corresponding $R$-band weight-map, from which the variation of the
response as a function of right ascension can be seen.  Unfortunately,
poor weather conditions early in the observing period limited the
optical observations of the HDF1 (STIS) field to the $R$-band and
about one-third of the desired exposure in \i-band. This implies that
the easternmost edge of the WFPC2 field lies in the region, some 30
arcsec wide, where the sensitivity of the SUSI2 exposures drops
sharply.

Infrared observations were obtained using the SOFI camera (Moorwood,
Cuby \& Lidman 1998) also at the NTT. SOFI is equipped with a Rockwell
1024$^2$ detector that, when used together with its large field
objective, provides images with a pixel scale of 0.29 arcsec, and a
field of view of about $4.9 \times 4.9$ square arcmin. Because of the
somewhat smaller field of view of SOFI, the infrared pointings were
chosen to have the fully sampled pixels (discarding the edges due to
the jitter pattern) overlapping and covering both STIS and WFPC2.
This led to the centers listed in Table~\ref{tab:pointings} and the
coverage illustrated in Figure~\ref{fig:hdfs}. The final areas covered
by SOFI are 21.3 and 21.5 square arcmin in HFD1 and HDF2,
respectively, with some overlap.

The infrared observations were jittered relative to the centers given
above. The procedure consists of a series of short exposures with
small position offsets from the target position. This strategy is used
to deal with a background that (i) will saturate the chip in a short
time ($\sim 1$ minute), and (ii) varies on short time-scales. The aim
is to remove the sky signal in each pixel of the image, using
observations of the same pixel as it points at different parts of the
sky.  The jitter strategy has been implemented as a standard observing
template ({\em AutoJitter}) for the SOFI instrument.  Using this
template, offsets are generated randomly within a square box of a
specified size $s$, chosen to be $s=45$ arcsec, approximately 10\% of
the SOFI detector field of view. These offsets are constrained so that
all distances between pointings, in a series of 15 consecutive
pointings are larger than 9 arcsec. Individual observations comprised
sixty one-minute exposures with offsets generated by the AutoJitter
template. Each exposure consisted of the average of six ten-second
sub-exposures.

\begin{figure*}
\resizebox{0.95\textwidth}{!}{\includegraphics{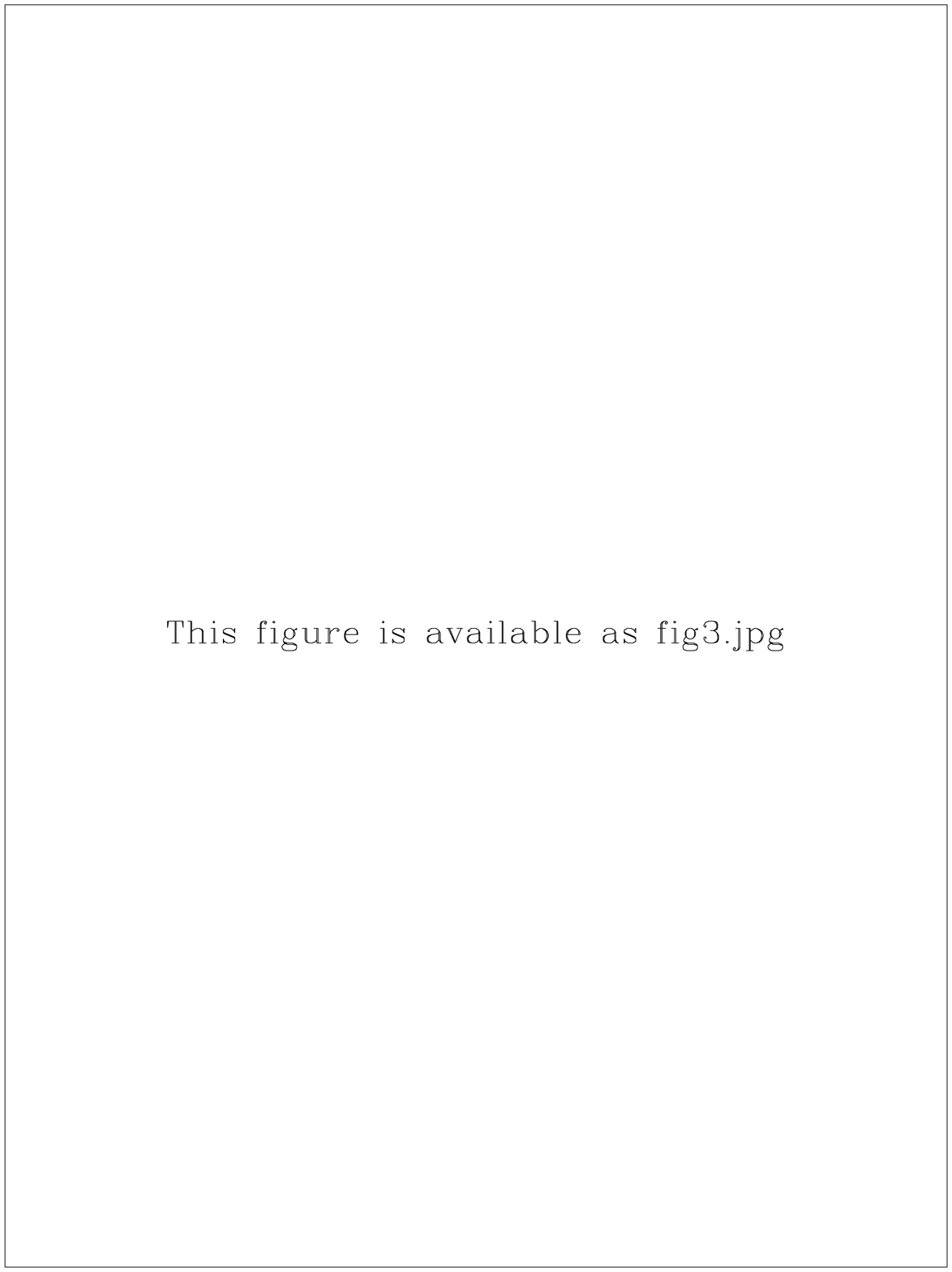}}
\caption{Final HDF2 co-added image for each of the optical passbands
(from left to right, top to bottom: U, B, V, R, and I). In the red
passbands one sees the light streaks from bright stars outside the
field and in the blue ($U$) from the bright star in the upper right
(northwest) corner of the images.}
\label{fig:opmosaic}
\end{figure*}

\subsection{Optical Data}

The optical observations were carried out in the period
August-November 1998, using broad-band $UBVRI$ filters (ESO \# 810,
811, 812, 813, and 814, see SUSI2 web page). A total of 12 nights were
allocated for optical EIS-DEEP observations of the HDF-S and AXAF
fields. However, because of poor weather conditions it was not
possible to complete the program as originally envisioned. While the
optical observations of HDF2 (WFPC2) field are complete, only $R$ and
some $I$-band data are currently available for HDF1 (STIS) and no
observations were made of the NIC3 field.

Table~\ref{tab:exposures} summarizes the observations, listing for
each field and passband the total integration time, the number of
exposures, the range of seeing as measured on individual exposures,
the full-width at half-maximum (FWHM) of the point spread function
(PSF) in the final co-added image, and the estimated $1\sigma$
limiting isophote within a 1 square arcsec area. While most of the
observations were conducted in photometric nights with relatively good
seeing, data were also taken in nights when the transparency and
seeing varied considerably.  Single exposures ranged from 800 sec
($U$) to 200 sec ($I$).

Landolt standards taken from Landolt (1992) were observed during all
clear nights and in all passbands used, thus allowing an accurate
determination of the photometric zero-point, extinction and color
terms. Independent solutions were found for each run typically using
some 50 standard star measurements.  A total of 9 nights were deemed
photometric and frames in all passbands are available in these nights,
thus allowing for a suitable absolute calibration of the
observations. The photometric calibration was done independently for
the two chips of the camera. Magnitudes for Landolt stars were
obtained using an aperture  14 arcsec in diameter.  Comparison of
the listed magnitudes of the Landolt standards with the derived
photometric solutions yields the following estimates for the accuracy
of the zero-points: $\pm$ 0.1~mag in \u; $\pm$0.03~mag in \b;
$\pm$0.03~mag in \v; $\pm$0.02~mag in \r; and $\pm$0.05~mag in
\i. These results apply to both chips, with the relative zero-point
difference between the chips being smaller than their estimated
errors. Relative to the Johnson-Cousins (JC) the following color terms
have been computed: 

\medskip
$
\begin{array}{l}
(U_{JC}-U_{EIS})= 0.14 (\pm 0.06) \times(U-V)_{JC},\\
(B_{JC}-B_{EIS})= 0.14 (\pm 0.02) \times(B-V)_{JC},\\
(V_{JC}-V_{EIS})= 0.0 (\pm 0.01) \times(B-V)_{JC},\\
(R_{JC}-R_{EIS})= -0.04 (\pm 0.01) \times(V-R)_{JC},\\
(I_{JC}-I_{EIS})= 0.03(\pm 0.02) \times(V-I)_{JC},\\
\end{array}
$

The magnitudes were also corrected for galactic absorption, using
$E(B-V)=0.027$ as derived from Schlegel, Finkbeiner \& Davis (1998),
yielding $A_U=0.14$~mag, $A_B$=0.08~mag, $A_V$=0.07~mag,
$A_R$=0.05~mag and $A_I$=0.04~mag.  To facilitate the comparison with
the HDF data and other authors, below all magnitudes given below,
unless otherwise specified, have been converted to the $AB$ system
using the following relations: $U_{AB} = U + 0.82; B_{AB}= B - 0.06;
V_{AB}= V; R_{AB}= R + 0.17;$ and $I_{AB}=I + 0.42$.

A total of 564 standard stars and 226 science frames were reduced
using standard IRAF tasks. For each frame a pre-scan correction was
applied, using the pre-scan region associated to each chip, and the
frames trimmed.  A master bias for each run was created by median
combining all bias frames and applying a 3-$\sigma$ clipping. The same
procedure was adopted for the dome flats and sky-flats. All frames
were corrected using the dome flat and an illumination correction
based on the sky-flats. For the red passbands, in particular $I$, the
images exhibit significant fringing. To partly remove the observed
low-level pattern ($\lsim$5\%) background maps were produced using
SExtractor and were combined to create a superflat. The frames were
then flatfield corrected using this superflat and each image was input
to the IRAF {\it mkfringecor} task. The output fringe images were then
combined into a master fringe frame.  Finally,  the master fringe
frame was subtracted from the  science frames. It should noted that
this procedure may result in the loss of faint objects. However,
visual inspection of the fringe corrected images shows that the
improvement of the images outweighs the loss of faint objects.

After all these corrections were applied, several hot pixels were
still visible in the individual $U$ and $B$-band exposures, especially
in the CCD \#46. While some were isolated pixels, others were
clustered in four different regions forming complex patterns. These
patterns were clearly visible after the images in a given band were
co-added, leading to a significant number of spurious detections. To
minimize their effect, a suitable mask was created and used in the
co-addition. To create this mask the following procedure was
adopted. All images in $U$ and $B$ were median-combined and a high
count threshold was set so that only the hot pixels were visible,
thereby allowing them to be masked out. Finally, after the frames were
corrected for all instrumental effects, an eye-inspection was carried
out to mask out other features such as satellite tracks and reflection
spikes associated with two relatively bright stars outside and one
inside the field covering the WFPC2 camera. 

\begin{table}
\caption{Summary of Infrared  Observations}
\begin{tabular}{lrcccc}
\hline\hline
Filter & $t_{total}$   & $N_f$ &seeing range & FWHM &   $ \mu_{lim}$\\
       &    (sec)    &        &  (arcsec)    & (arcsec) & (mag arcsec$^{-2}$)\\
\hline
HDF1  &      &             &      &   & \\
\hline
  \j & 10800 & 180  & 1.3-1.4 & 1.37 &   25.37  \\
  \h & 3600 &  60  & 0.9-1.0 & 0.91 & 23.78  \\
  \k & 10800 & 180 & 0.9-1.0 & 0.90 & 23.83 \\
\hline
HDF2  &      &            &       &   & \\
\hline
   \j & 10800 & 180 &0.9-1.0  &  0.90 &  25.94 \\
   \h & 7200 &  120 &0.8-0.9 & 0.85&  24.06 \\
   \k & 18000 & 300 &0.7-1.2 & 0.96 & 24.22 \\
\hline\hline
\end{tabular}
\label{tab:obsir}
\end{table}

\subsection{Infrared Data}

Infrared observations in the $JHKs$ bands, for which a total of 10
nights were allocated, were obtained during the same period as the
optical data. Total integration times, the number of frames, the
seeing range, the FWHM measured on the co-added images, and the
estimated $1\sigma$ limiting isophote within one square arcsec are
given in Table~\ref{tab:obsir}.

During all nights infrared standards taken from Persson (1997) were
observed. From the photometric solutions the errors in the absolute
photometric zero-points are found to be: $\pm$ 0.1~mag in \j; $\pm$
0.05~mag in \h; and $\pm$ 0.1~mag in \k\ for HDF1; and $\pm$ 0.05~mag
in \j; $\pm$ 0.05~mag in \h; and $\pm$ 0.05~mag in \k\ for HDF2.
These magnitudes were also converted to the $AB$ system, using:
$J_{AB}= J + 0.89; H_{AB}=H+1.38;$ and $K_{AB}= Ks + 1.86$.

The reduction of SOFI imaging data which include a total of 545
standard stars and 1020 science frames requires one major step
particular to infrared data. This step uses the individual exposures
to filter and subtract the sky-background. Because the telescope
pointing error introduces an uncertainty in the image offsets, the
exact offsets have to be determined, so that all exposures can be
shifted and added. All these steps have been incorporated in the
program {\em jitter}, from the astronomical data-reduction package
{\em eclipse}, written by N. Devillard (Devillard, 1998).

The background filtering and subtraction uses the algorithm {\em
sky-combine} in the {\em jitter} program. This algorithm finds, for
each pixel in an exposure, the 14 pixel values for the 7 observations
prior to and subsequent to that observation. Each of the 15 pixel
values (including the current pixel) is divided by the median sky for
the exposure it belongs to. The two smallest and largest values are
rejected from the scaled array of 15 pixels, and the remaining values
are averaged and subtracted from the central pixel.  After sky
subtraction, the image offsets between each exposure and the first
exposure, which has by definition a zero offset from the target, are
determined. In the first exposure a region of interest is defined by
choosing an area of $25\times 25$ pixels around a bright peak near the
center of the image. For each subsequent image the telescope offset is
determined by cross-correlating the region of interest in the first
image with the equivalent region in the offsetted image. Finally, all
images are re-sampled into an output image for which the projection is
equal to the projection of the first image, using a hyperbolic tangent
kernel.  After all exposures have been re-sampled, they are averaged
to produce one final de-jittered image. The {\em jitter}-program also
takes care of bias subtraction, flatfielding and interpolation of
bad-pixels prior to sky subtraction, as well as cleaning of columns
that are affected by ``bleeding'' from very bright stars in the
combined image after the shift and add procedure. Because of the
jitter pattern the sensitivity of the combined image falls of at the
edges. Therefore, the effective area is $4.6 \times 4.6$ square
arcmin.

\subsection{Processing}

After removing the instrumental signatures, both optical and infrared
images were input to the EIS pipeline for astrometric calibration
using the USNO-A V1.0 catalog as reference. In the case of SUSI2,
independent astrometric solutions were found for the two chips.
Before the astrometric calibration, images were processed by
SExtractor using a high detection threshold to measure the size of the
PSF of each frame, to create weight maps for co-addition and to flag
cosmic rays and defects using the Artificial Retina algorithm
described in Nonino
\etal (1998).  These features were incorporated into the mask created
earlier based on the hot pixels, bad columns and diffraction
spikes. Images in the same passband were then co-added using the
"drizzle" method, originally created to handle HST images (Fruchter \&
Hook 1998), at the same resolution as the original images. In the
process of co-addition, images taken in photometric nights were used
as reference and all other images were corrected to have the same
instrumental magnitudes after extinction correction. Finally, the
absolute zero-point was determined and stored in the header of the
co-added image.

\begin{figure*}
\resizebox{0.95\textwidth}{!}{\includegraphics{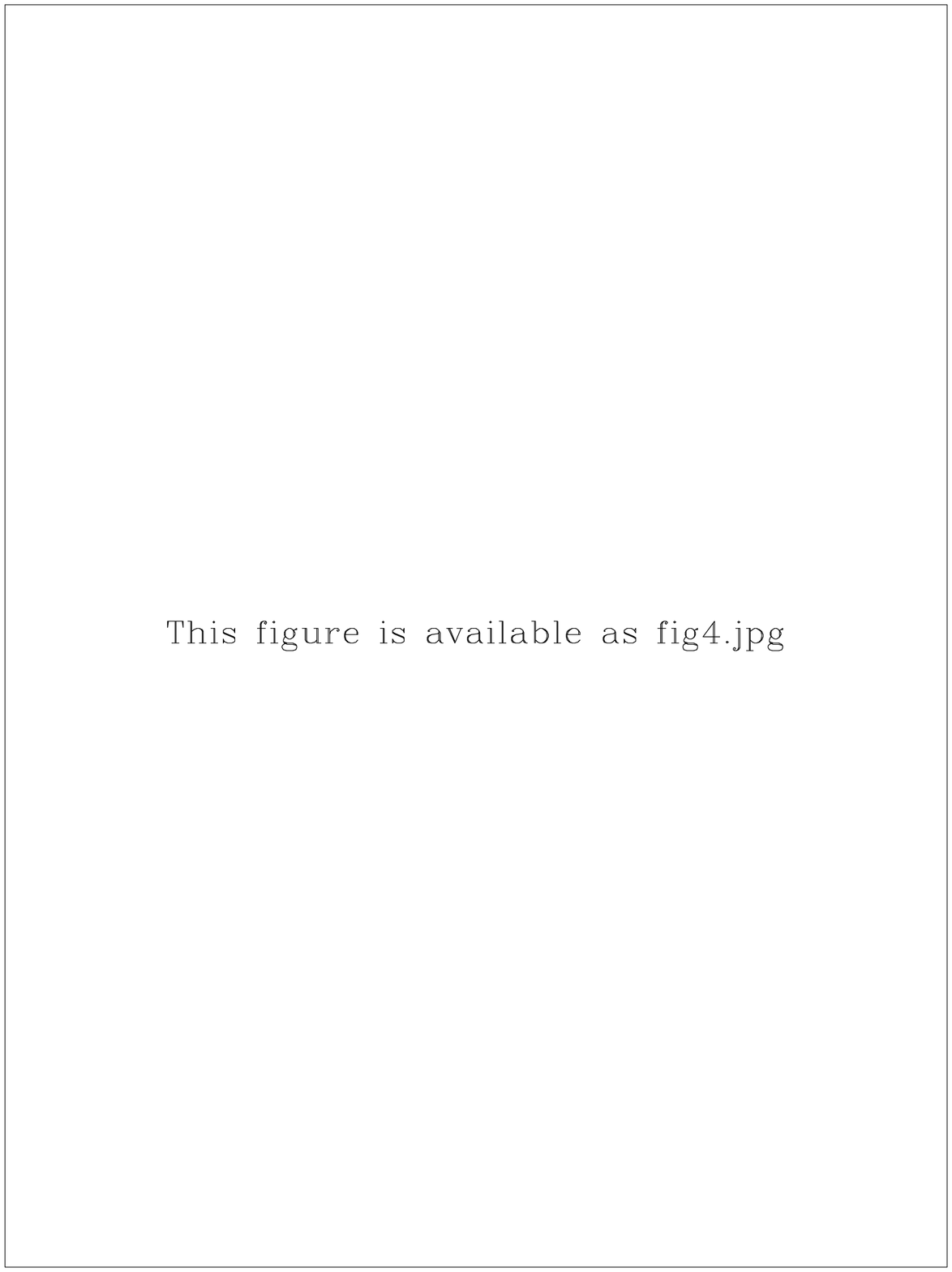}}
\caption{Final co-added infrared images in the passbands \j\ (upper
panel), \h\ (middle panel) and \k\ (lower panel). These observations
cover both HST-WPFC2 and STIS fields.  Note that the integration time
in \h-band is less than the others and that the \k\ image over the
WFPC2 field is considerably deeper. The images are about $9.0 \times
4.8$ square arcmin. }
\label{fig:irmosaic}
\end{figure*}

\begin{figure}
\resizebox{9cm}{!}{\includegraphics{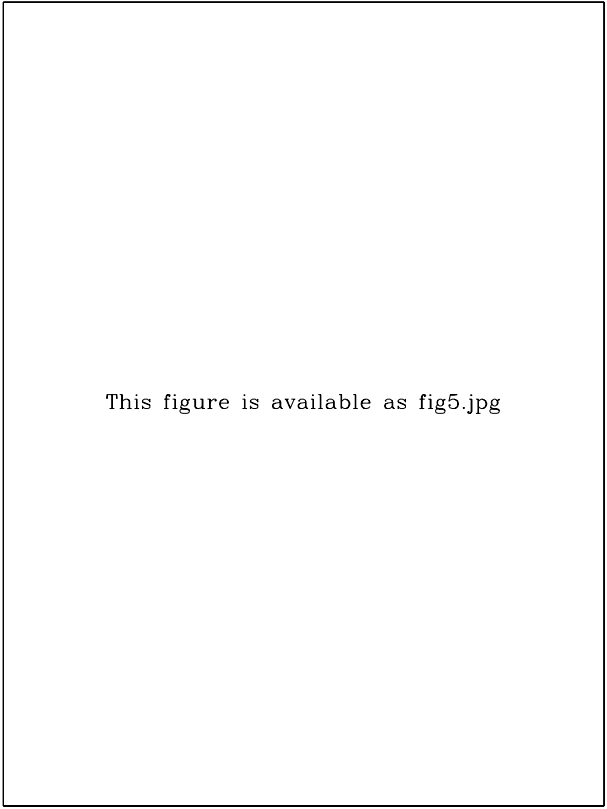}}
\caption{Comparison between the \u-band images from WFPC2 and
SUSI2. While the resolution from space is unmatched, the detection
limit is comparable even though the exposure time of the ground-based
observations is 1/10 of that from HST. However, it
should be noted that the passbands are different with that of WFPC2
located blueward compared to that of SUSI2.}
\label{fig:u}
\end{figure}

The resulting co-added optical images for the HDF2 field in each
passband are shown in Figure~\ref{fig:opmosaic}.  These are the full
images without any trimming of the edges where the dithering pattern
can be seen.  Note that the \r-band image of HDF1, covering the
HST-STIS field shows some stray light and the one hour \i-band data
have not yet been fully reduced.  Once completed they will immediately
be made publicly available. From Figure~\ref{fig:opmosaic} one can
immediately see why this region is not appropriate for ground-based
deep imaging. In all passbands the influence of the bright star on the
upper right (northwest) can be clearly seen.  In \u\ it shows as a
bright halo which is not appropriately modeled by the object detection
algorithm, leading to an increase in the number of detections in this
region. The effects of other stars just outside the field of view are
also clearly seen in the top part of the images in all in bands,
becoming particularly strong in the red passbands.  Unfortunately,
these stars could not be avoided given the other constraints on the
pointing (see Section~\ref{obs}). It is also worth emphasizing the
superb performance of SUSI2 in the blue passbands. Even though the
SUSI2 images cannot, of course, match the resolution of WFPC2 the
limiting magnitude is comparable to that achieved from space in 1/10
of the time. This can be seen in Figure~\ref{fig:u} where a small
region of the sky observed by the HST-WFPC2 and SUSI2 are
compared. Note, however, that the response of the WFPC2 FW300 filter
peaks further in the $UV$.

Similarly, co-added images were produced for the infrared observations
for each band and pointing. However, since in this case there are
observations of two adjacent fields, which overlap, additional
co-added images, covering the WFPC2 and STIS fields, have been
produced for each of the available passbands. These are shown in
Figure~\ref{fig:irmosaic}. Note that the depth of the images may vary
because of the different integration times in different fields. These
infrared data which offer a unique complement to the HST observations
are being made public world-wide.

\begin{figure*}
\resizebox{9cm}{!}{\includegraphics{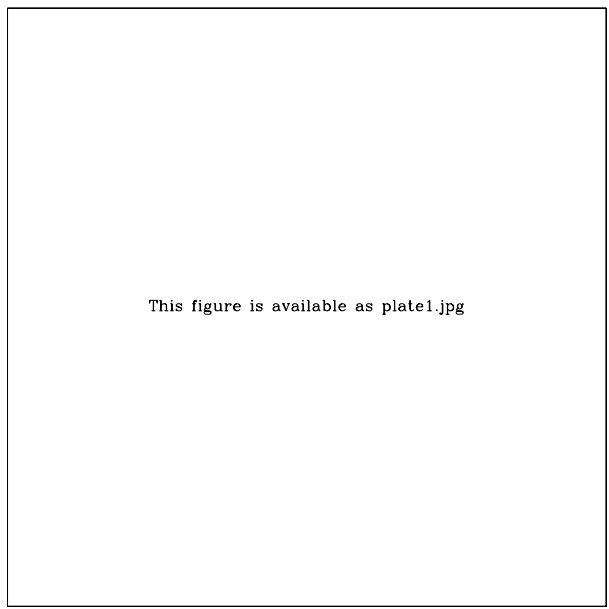}}
\caption{True-color image of the HDF2
field based on the five observed optical passbands. The blue channel
is represented by the \u\b-band images, the green channel by the
\v\r-band images, and the red channel by the
\i-band image. The edges of the field have been trimmed to include
only pixels with a sensitivity $\gsim$ 70\% in all passbands. }
\label{fig:plate1}
\end{figure*}

Figure~\ref{fig:plate1} is a true-color image of the HDF2 field with
the blue channel represented by the \u+\b-band images, the green
channel by the \v+\r-band images, and the red channel by the \i-band
image. As expected the image is dominated by faint blue galaxies,
however several examples of red objects, some of which possibly high-z
candidates can also be found.

\begin{figure*} 
\resizebox{9cm}{!}{\includegraphics{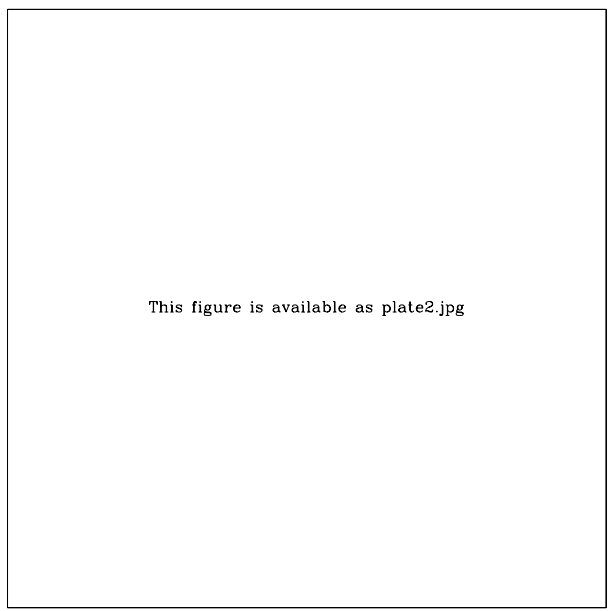}}
\caption{True-color image of part of  the HDF2 field for which the
optical and infrared observations overlap.  The region includes most
of the HST-WFPC2 field. The image is, perhaps, one of the most
colorful pictures available, being based on 8 passbands covering an
extended spectral region from 300 to 2200 nm. The blue channel is
represented by the combination of the \u\b\v-band images, the green
channel by the
\r\i-band, and the red channel by the \j\h\k-band images. This color
image covers an area of approximately $2.5\times4.0$ square arcmin
corresponding to the SUSI2-SOFI overlap covering the HST-WFPC2 field. }
\label{fig:plate2} \end{figure*}

In addition, in order to allow the production of multi-color catalogs based
on a single reference frame for detection, a co-added image of the optical
data matching the infrared images resolution has also been created.
Figure~\ref{fig:plate2} shows a true-color image of the HDF2
field with the blue channel represented by the \u+\b+\v-band images, the
green channel by the \r+\i-band images, and the red channel by the
\j+\h+\k-band images. This color image covers an area of approximately
$2.5\times4.0$ square arcmin, corresponding to the SUSI2-SOFI overlap
covering the HST WFPC2 field.

All the co-added images, including the weights and masks, are public
and may be requested from the URL "http://www.eso.org/eis".

\begin{table*}[tp]
\caption{HDF-South. \u-band Catalog of Sources in the HDF2 field.}
\label{tab:uband}
\resizebox{\textwidth}{!}{\includegraphics{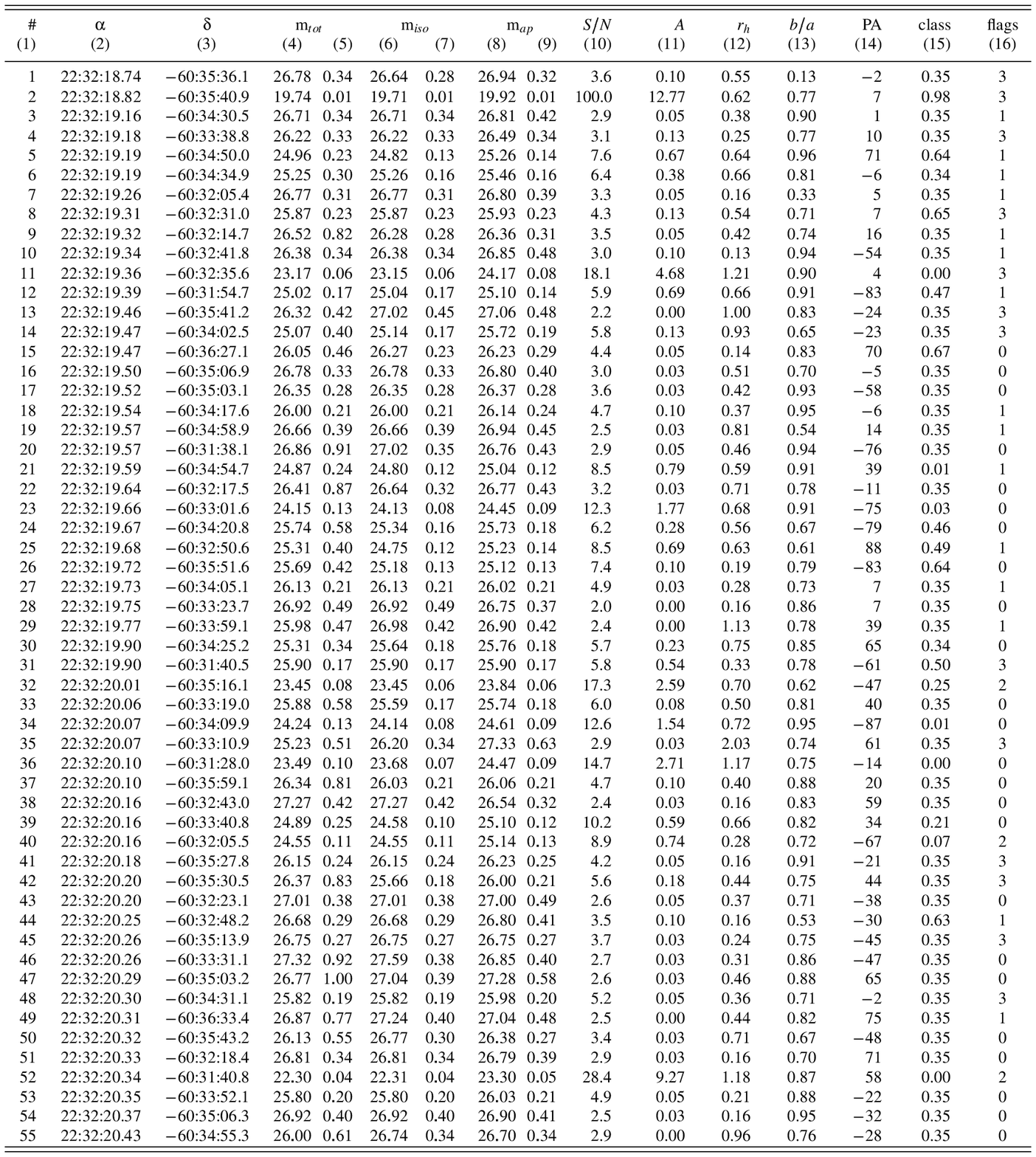}}
\end{table*}

\begin{table*}[tp]
\caption{HDF-South. \k-band Catalog of Sources in the HDF2 field.}
\label{tab:kband}
\resizebox{\textwidth}{!}{\includegraphics{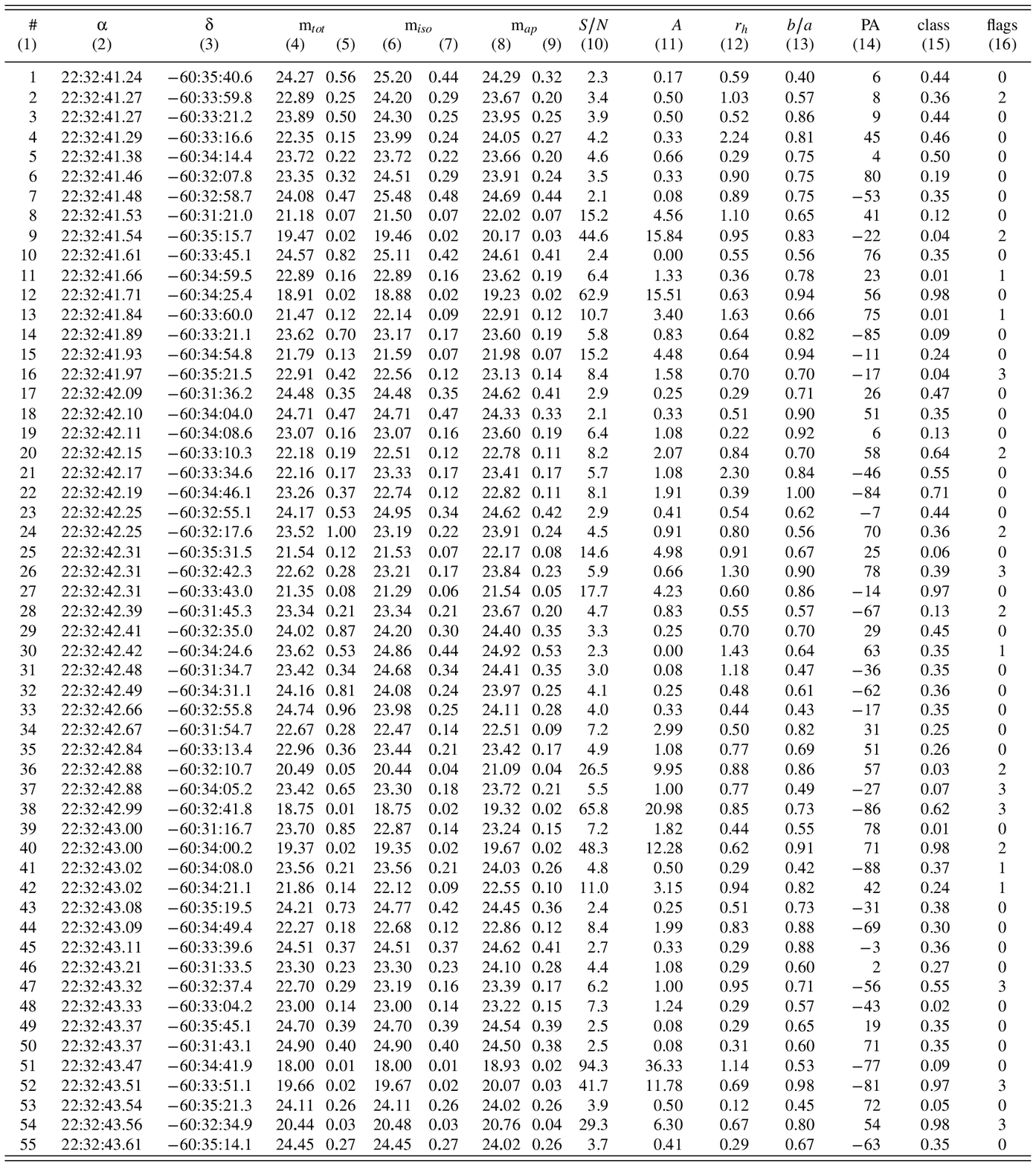}}
\end{table*}

\begin{table*}[tp] 
\caption{HDF-South. Multi-Color Optical Catalog from the $\chi^2$
image in the HDF2 field.}
\label{tab:chi2}
\resizebox{\textwidth}{!}{\includegraphics{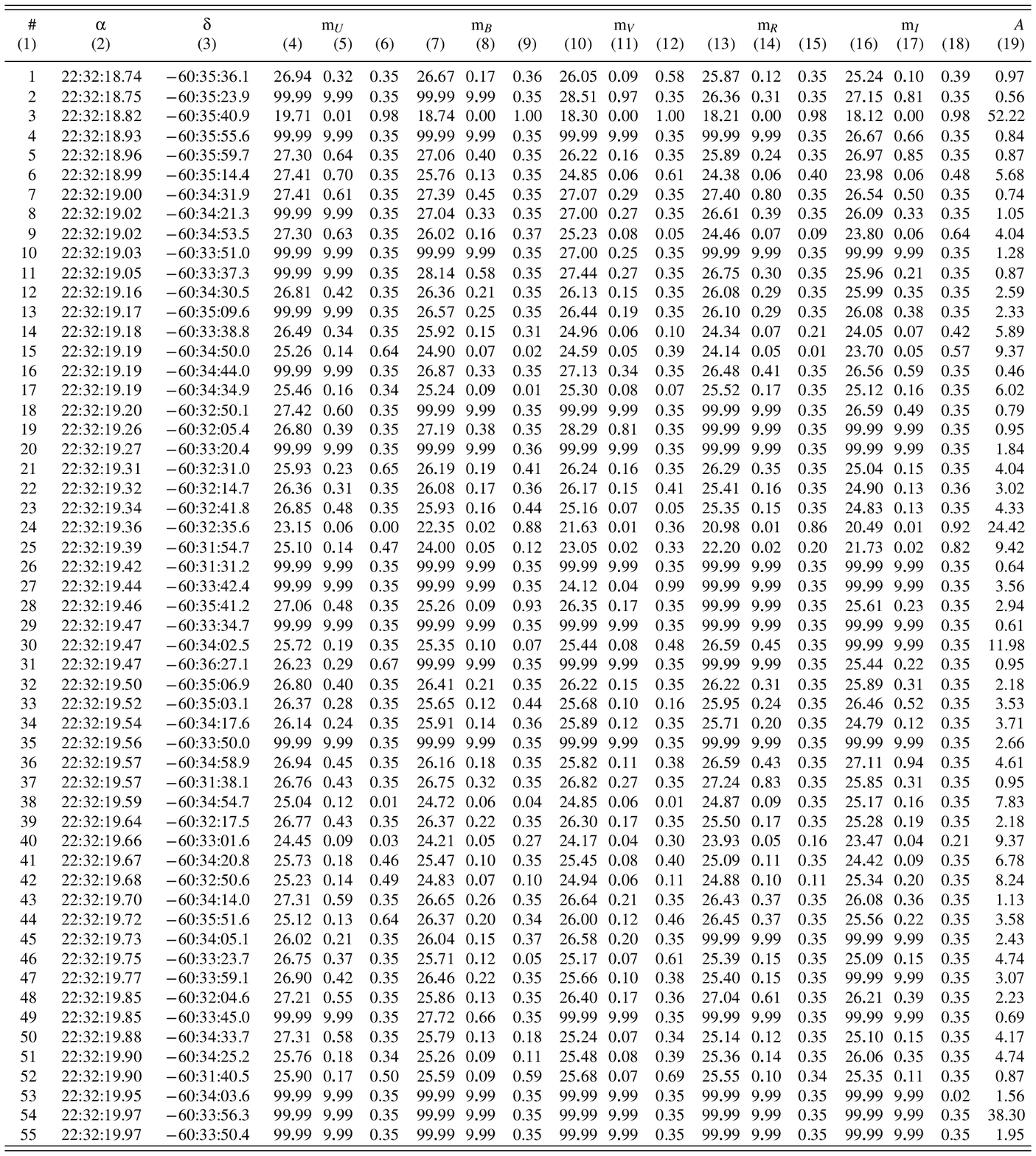}}
\end{table*}

\begin{table*}[tp] 
\caption{HDF-South. Multi-Color Infrared Catalog from the $\chi^2$ image in the HDF2 field.}
\label{tab:irchi2}
\resizebox{\textwidth}{!}{\includegraphics{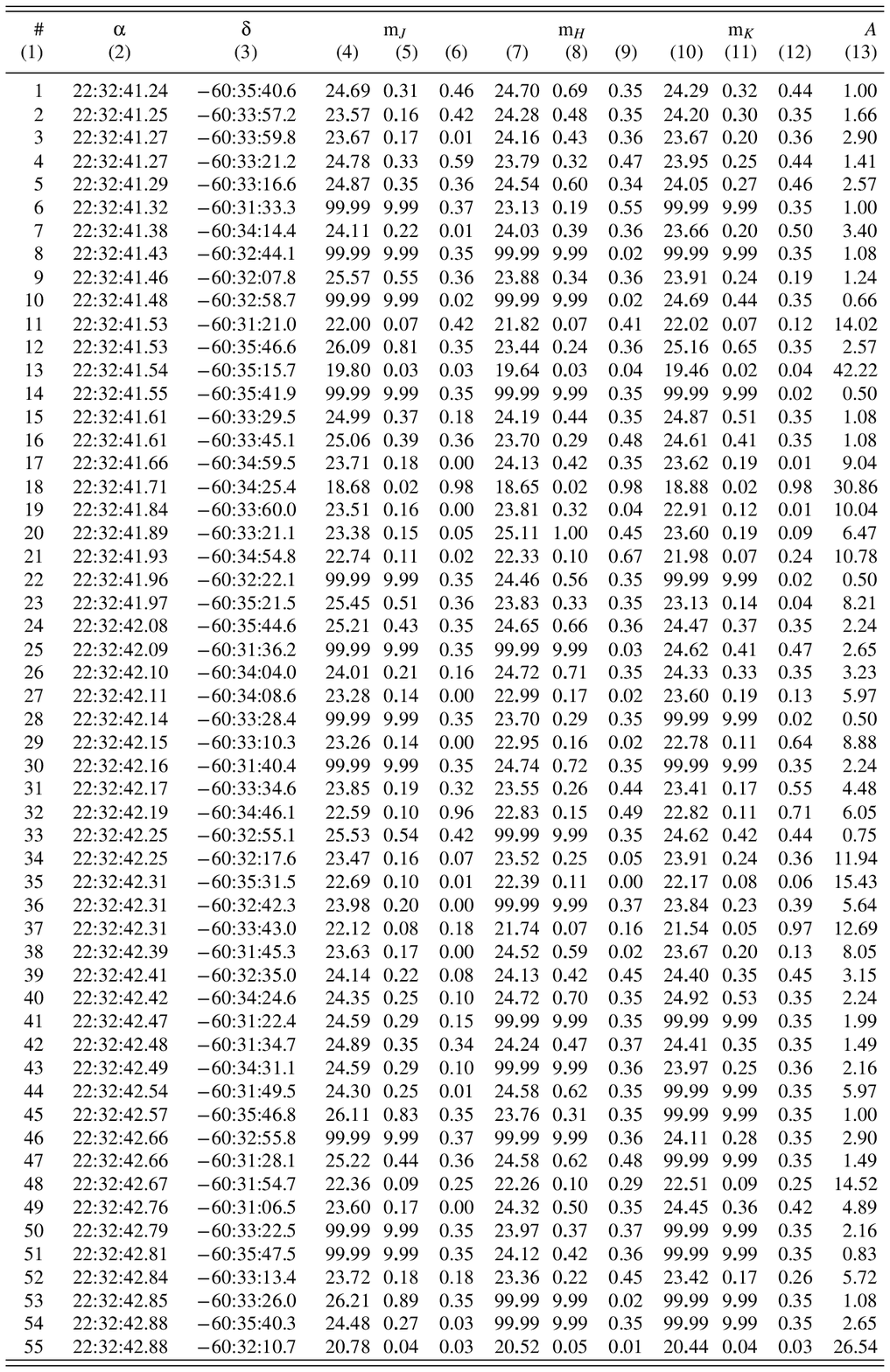}}
\end{table*}

\begin{table*}[tp] 
\caption{HDF-South. Multi-Color Optical-Infrared Catalog from the
$\chi^2$ image in the HDF2 field.}
\label{tab:optirchi2}
\resizebox{0.90\textwidth}{!}{\includegraphics{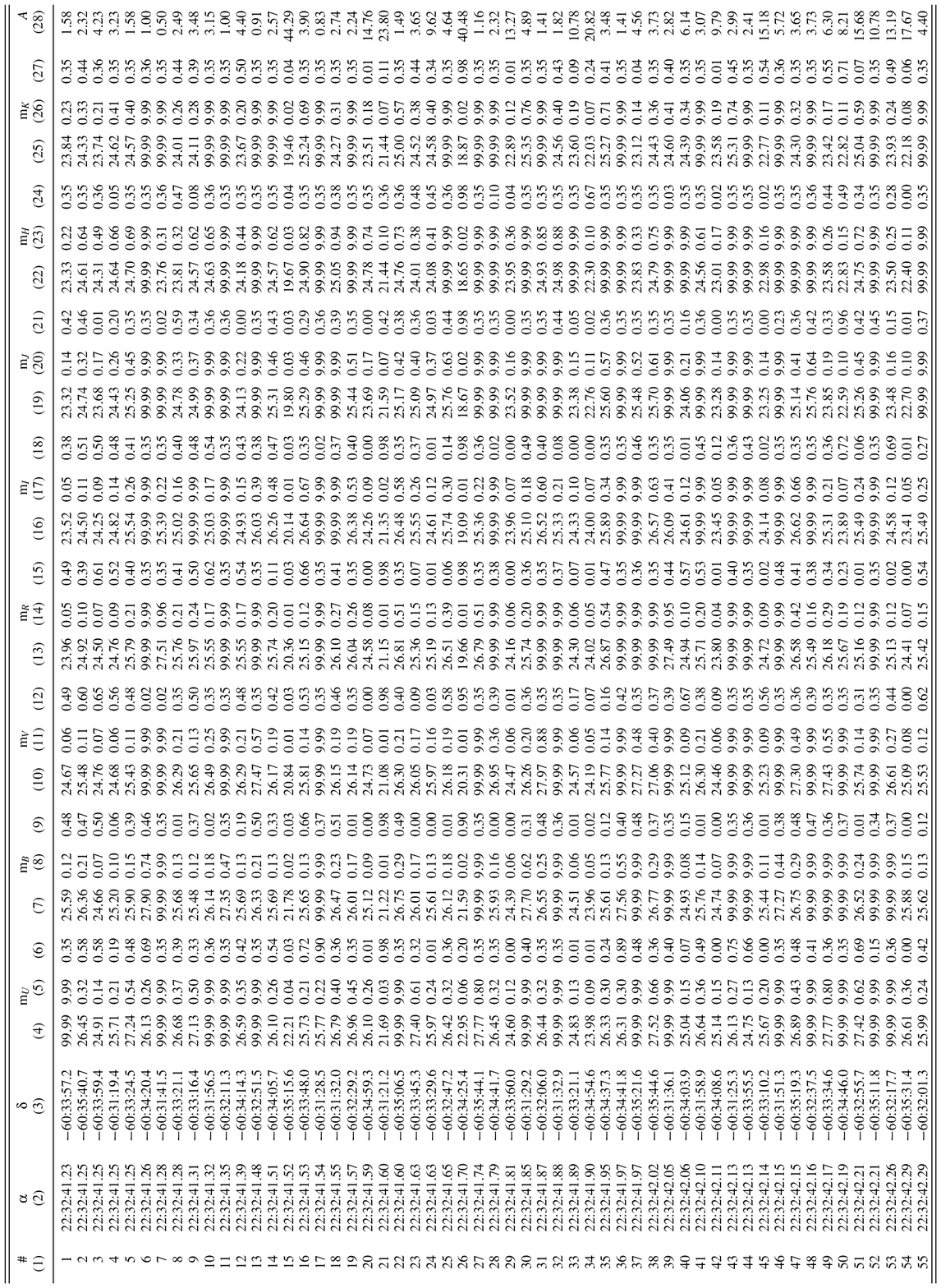}}
\end{table*}

\section {Object Catalogs}
\label{cats}

The recent availability of large samples of high-quality, digital
multi-band data, and improved techniques for assigning redshifts based
on photometric data, have motivated great interest in the production
of color catalogs of faint galaxies. Even though most methods use the
same basic steps, the differ in the details on how to detect and
measure magnitudes and colors as the detection threshold approaches
the noise level. In general, when dealing with multi-band data, color
catalogs can be produced either by the association of objects detected
independently in each passband or by using some reference image to
carry out the detection and to define apertures to be used for the
subsequent measurement of magnitudes in each passband. Again the
method of choice depends on the specific scientific goals, with each
method susceptible to different bias. To avoid some of these issues
single passband catalogs have been produced as well as examples of
multicolor catalogs produced using a reference frame. A complete list
of the catalogs currently available can be found on the EIS web page.

\subsection {Single Band Catalogs}

Source extraction was performed using the SExtractor software (Bertin
\& Arnouts 1996). Detection was carried out separately using the
co-added image of each passband and field. Star/galaxy classification
was based on the stellarity-index given by SExtractor, which is
approximately the probability of an object being a point-source.  The
main parameters in the detection are the smoothing kernel, taken to be
a Gaussian with a FWHM comparable to that of the PSF as measured on
the frame; the SExtractor detection threshold, taken to be 0.7; and
the minimum number of pixels above the detection threshold, taken to
be 10 and 7 pixels for the SUSI2 and SOFI images, respectively,
because of the different pixel scales. As an illustration, the listing
of the first 55 entries in the HDF2 source catalogs for \u\ and
\k-bands are presented in Tables~\ref{tab:uband}-~\ref{tab:kband}. 
For each object the following parameters are given:

Column (1) - the entry number in the table (not to be misinterpreted as
a unique reference to a specific object);

Columns (2) and (3): right ascension and declination (J2000.0);

\begin{figure}[t]
\resizebox{0.45\textwidth}{!}{\includegraphics{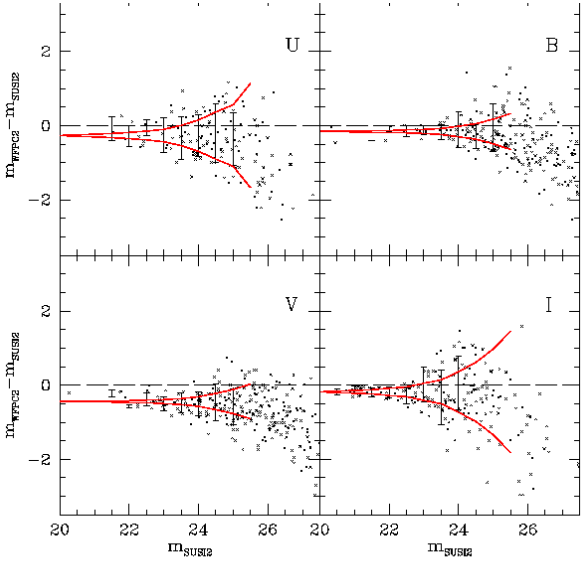}}
\caption{Comparison between total magnitudes computed from the
 WFPC2 image, convolved with a Gaussian of FWHM of 1 arcsec, with
those measured in the SUSI2 fields for the different passbands
indicated in each panel.}
\label{fig:wfpc2}
\end{figure}

Columns (4)-(9): total, isophotal (as measured by SExtractor) and
aperture (2 arcsec in diameter) magnitudes and respective errors. The
magnitudes have been corrected for extinction and converted to the
$AB$ system. The errors are the estimates from SExtractor, which only
include the shot-noise on the measured source and background
counts. These errors should be considered as lower limits as they do
not include the contribution from random or systematic errors in
determining the mean sky level.  However, as shown in Nonino \etal
(1998) the SExtractor estimates are, in general, a good estimate of
the error except, perhaps, at the very faint limit (see
Figure~\ref{fig:wfpc2}). Also note that the rms of the background in
the co-added image is artificially small due to the effective
small-scale smoothing introduced by the drizzling process.  To correct
for this effect the error estimated by SExtractor must be multiplied
by a factor of 1.5.  This correction has been taken into account in
all the errors listed in all the object catalogs. Only objects
detected with signal to noise $S/N\geq2$ (based on the isophotal
magnitude) are included.  Note that for faint objects it may occur
that the total and aperture magnitudes have large errors.  For those
with magnitude errors above 1~mag these magnitudes are replaced with
the isophotal magnitude;

Column (10): an estimate of the $S/N$ of the detection, using the
errors estimated above for the isophotal magnitude;

Columns (11): the total area $A$ of the object in square arcsec;

Column (12): the half-light radius $r_h$  in arcsec;

Column (13) and (14): minor to major-axis  ratio and the position
angle;

Column (15): the stellarity index computed by SExtractor in the given
band. Note that faint objects have the same value of the stellarity
index ($\sim 0.35$) because they are beyond the classification limit
($I\sim22$, see below). However, at the faint limits of the catalogs
these are predominantly galaxies;

Column (16): SExtractor flags.

\begin{figure}
\resizebox{0.45\textwidth}{!}{\includegraphics{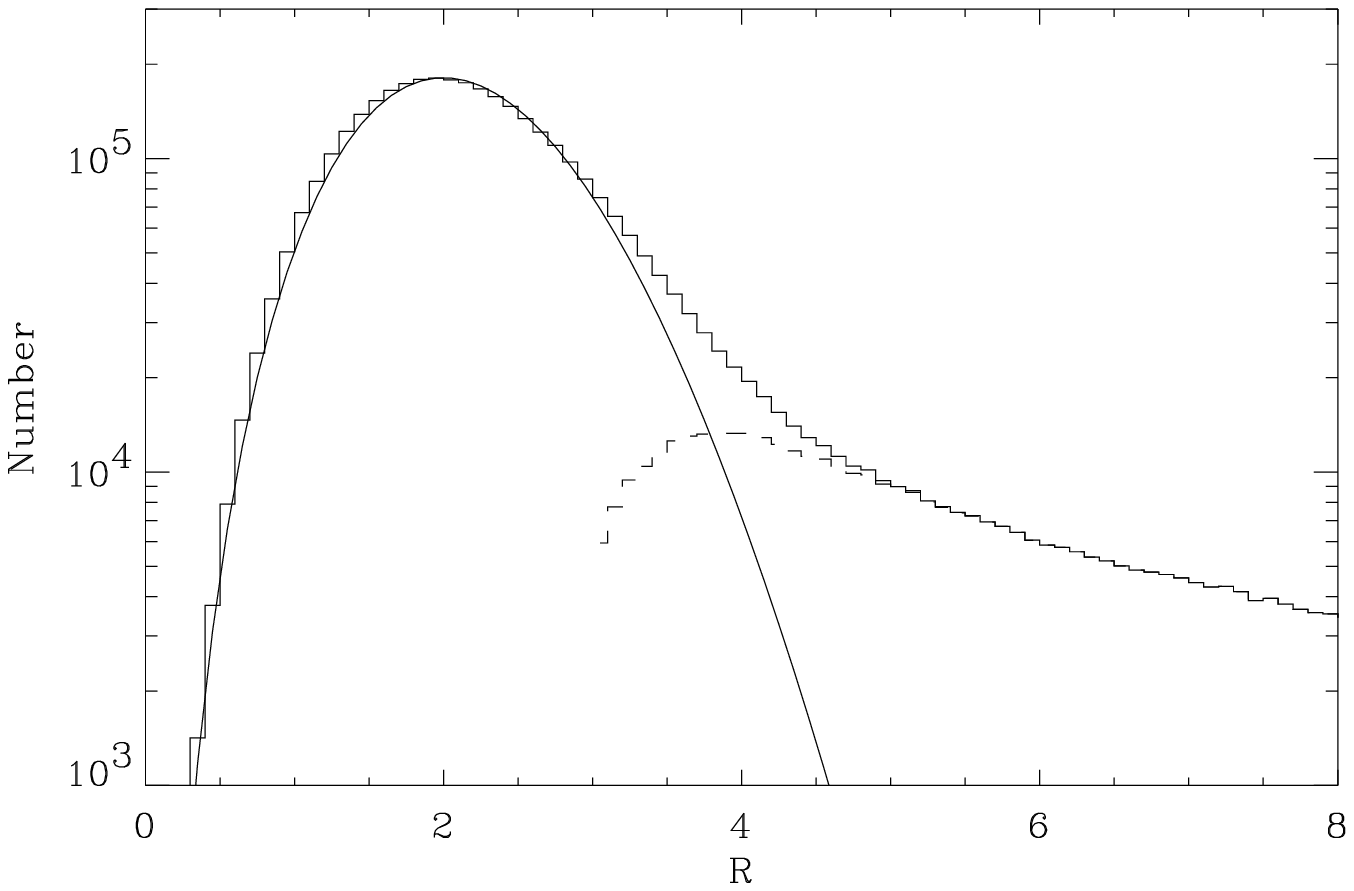}}
\caption{The $R$ distribution (solid histogram) for the $\chi^2$-image
produced by co-adding the optical data ($N=5$). This distribution is
compared to the expected $R$ distribution for pure Gaussian noise
(solid line), showing the excess tail at high $R$-values, due to
pixels containing object flux (dashed line).}
\label{fig:chi2dis}
\end{figure}

A total of five single optical-passband catalogs ($UBVRI$) are
available for HFD2 comprising 1227 objects in \u. 2018 in \b. 2071 in
\v, 1233  in \r\ and 1297 in \i.  Object catalogs have also been
extracted from the infrared images and Table~\ref{tab:kband}
illustrates the information available for single-band infrared
catalogs. It shows the first page of the deep \k-band catalog
extracted from the HDF2 images which overlap the HST-WFPC2 region.
The infrared catalogs comprise 905 objects in \j, 525 in \h\ and 638
in \k. For HDF1 the infrared catalogs have been constructed and
comprise of 415 objects in \j, 439 in \h\ and 594 in \k. Note that the
limiting magnitudes vary for different pointings and bands (see
Table~\ref{tab:obsir}).  The ASCII version of these catalogs are
available at ``http://www.eso.org/eis''.

\subsection{Multi-band Catalogs}

Using the single passband catalogs presented in the previous section a
multi-color catalog can be built by the association of sources
identified in the different passbands. An alternative is to carry out
the detection in some suitably chosen reference image and measure the
photometric quantities in each passband using the same circular or
isophotal aperture as measured in the reference image, thereby
assuring a one-to-one correspondence. In the past, different choices
have been made regarding the choice of this reference image (\eg
Williams \etal 1997, Arnouts \etal 1998, Szalay, Connoly \& Szokoly
1998). Without discussing the merits of each of these options, and
after some experimentation using either $R+I$ or \i\ as the reference
the $\chi^2$ method advocated by Szalay, Connoly \& Szokoly (1998) has
been adopted.

The method takes into account all the multicolor information
available. Since all photometric bands are registered to a common
reference system, each pixel has measured fluxes $f_i$, $i=1...N$,
with $N$ the number of photometric bands. ($N=3$ for SOFI data, $N=5$
for SUSI2 data, and $N=8$ for the combined optical and infrared
data). For each band the local mean background $\mu_i$ and its
standard deviation $\sigma_i$, have also been determined. Therefore,
each pixel can be described by an N-dimensional vector, whose length
$R$ is given by
\begin{equation}
R^2=\sum_{i=1}^{N}\frac{(f_i-\mu_i)^2}{\sigma_i^2}.
\end{equation}
The probability distribution for $R^2$ is then a $\chi^2$ with $N$,
degrees of freedom, or, for $R$
\begin{equation}
dP(R)=\frac{1}{2}R^{N-1}e^{-R^2/2}dR.
\end{equation}
Hence, by subtracting the backgrounds $\mu_i$, dividing by
$\sigma_i^2$, and subsequent co-addition of the squared images, a new
image of $R^2$ values is created. This $\chi^2$-image, measures the
cumulative probability $P(>R)$ that a pixel is drawn from the sky
distribution. In practice the square-root (e.g. $R$) image is
analyzed.

Figure~\ref{fig:chi2dis} shows the $R$ distribution for the
$\chi^2$-image that was produced by co-adding the optical data
($N=5$). This distribution is compared to the expected $R$
distribution, showing an excess tail at high $R$-values, due to
pixels containing object flux. The difference between the measured
distribution and the expected noise distribution gives the
distribution of $R$-values for pixels containing object flux, and the
ratio of the noise distribution and object distribution provides a
good measure of the probability that a given $R$ value is due to the
presence of an object. In principle, one could use the $R$-value for
which the number of object pixels exceeds the number of noise pixels,
as a threshold value for SExtractor, taking the $\chi^2$-image as
the detection image, and the single-band images as the analysis image.
However, in practice, one would like to take into account the
correlation of object flux over several pixels, to further enhance the
sensitivity to faint objects.

The procedure adopted for the present work was to convolve the
background subtracted co-added single band images with a Gaussian with
1 arcsec FWHM and compute its rms noise ($\sigma_i$). These convolved
images were then normalized by their respective rms maps, squared and
added to produce the $\chi^2$ image. In order to determine the
threshold and the minimum number of pixels to be used as a detection
criterion in SExtractor, empty images, containing randomly generated
noise, were created. In these random images the number of `false'
objects as a function of detection parameters was determined, thus
providing both a detection threshold and the minimum number of
contiguous pixels, which minimize the number of false
detections. These detection parameters were used to construct a first
set of catalogs for the science images. The detection threshold was
then lowered in several steps, until the number of additional sources
in the subsequent catalog became smaller than twice the number of
false sources that would be included by similarly lowering the
threshold in the random image. Hence, by design, at the lowest signal
to noise levels, the catalogs are roughly 50\% reliable.

This method was used to compile a multi-color optical catalog for
HDF2, comprising 2862 objects. This catalog gives for each object
detected on the $\chi^2$ image, built from the combination of the
$UBVRI$ co-added images, the following parameters. The first three
columns give the entry number in the table (not to be misinterpreted
as a unique reference to a specific object), the right ascension and
declination (J2000.0) as determined in the detection image.  These
columns are followed by three columns for each passband considered
(depending on the specific sample), listing the magnitude, the error
and the stellarity index. The final column gives the area of the
object as measured in the detection image in square arcsec. For large
objects, with a detection area larger than the aperture of 2 arcsec in
diameter, the magnitude is measured within the isophotal area.  For
smaller objects, the magnitude is measured in the 2 arcsec aperture.
For objects with $S/N\leq1$, the magnitude is set to 99.9 and the
error to 9.99 to indicate a detected object with insufficient flux in
the given band. In the Table~\ref{tab:chi2} only the first 55 entries
of this catalog are shown.

A discussion on the performance of this procedure for detecting
objects is discussed in Section~\ref{results}.  Two similar catalogs
are also available listing infrared colors for the two SOFI fields
comprising 1424 objects in HDF1 and 1530 in HDF2, respectively.
Table~\ref{tab:irchi2} lists the first 55 entries of the infrared
catalog produced for the HDF2 field. The format is similar to that
presented in the previous table. Finally, Table~\ref{tab:optirchi2}
lists the first 55 objects extracted from the combined
optical-infrared $\chi^2$ image.  The corresponding catalog contains
1202 objects detected within the HDF2 SOFI-SUSI2 overlap, which covers
the HST-WFPC2 field. Of the 1202 objects detected in the combined
$UBVRIJHKs$ catalog, 479 objects have measurable magnitudes in all the
individual passbands; 52 objects in none of the
individual passbands; and 61 in only one of the passbands. Most of the
latter 61 objects have a measurable magnitude in either the blue (\u or
\b) or red (\j or \k) passbands.

\begin{figure}[t]
\resizebox{0.40\textwidth}{!}{\includegraphics{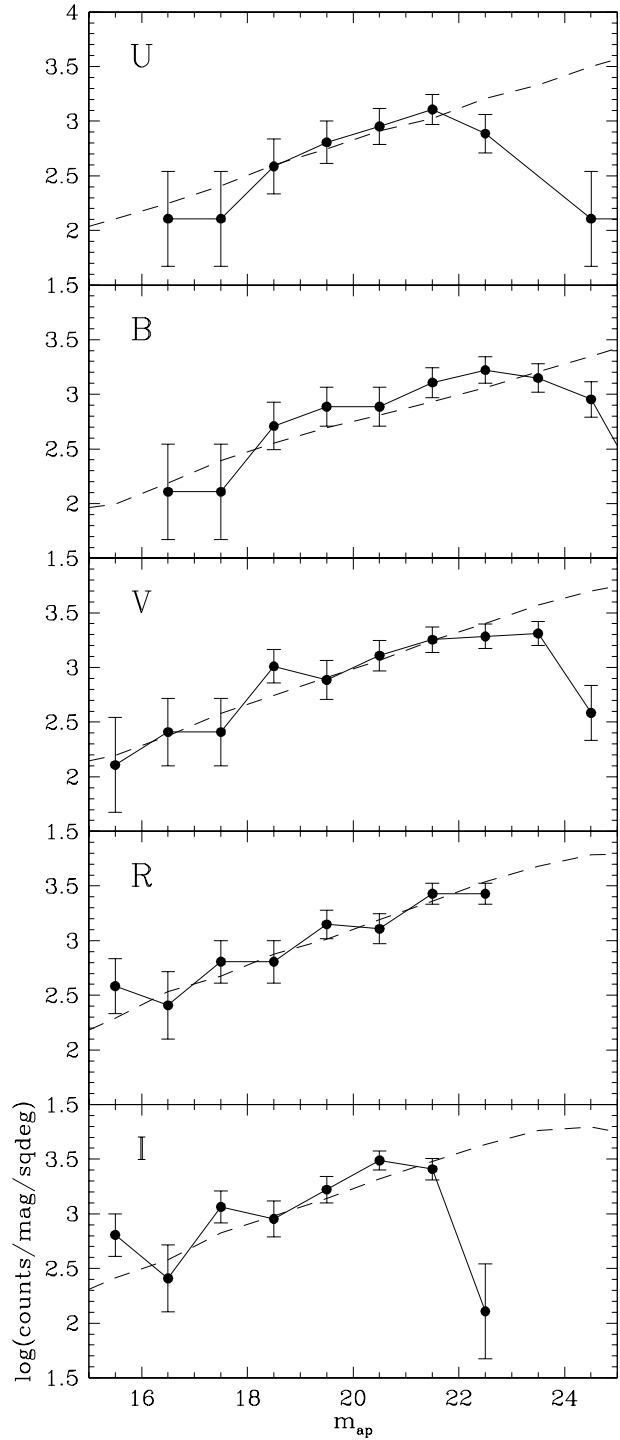}}
\caption{Comparison of the observed star-counts (filled circles) 
with model predictions as described in the text (dashed line).}
\label{fig:ncounts_star}
\end{figure}

\begin{figure}
\resizebox{0.43\textwidth}{!}{\includegraphics{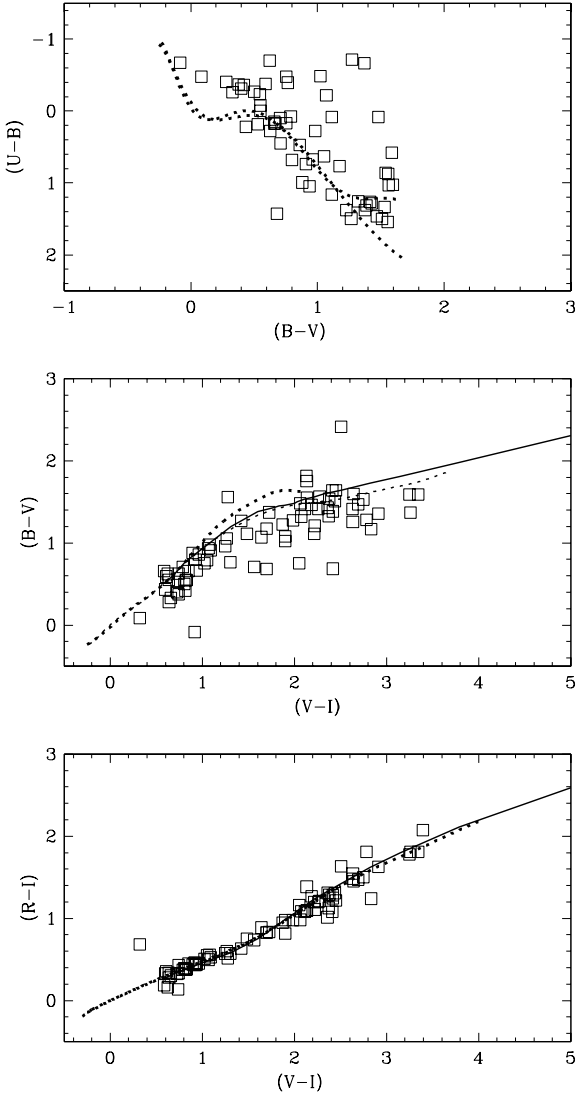}}
\caption{Color-color diagrams for point sources in the HDF2 field
compared with empirical determinations (dashed line) and model
predictions (solid line). Only $5\sigma$ detections are included and
the number of points shown are 58, 75 and 77 from the top to the
bottom panel.}
\label{fig:starcol}
\end{figure}

\section{External Comparison}
\label{comp}

To check the photometric accuracy of the EIS-DEEP data, these were
compared to other ground-based observations of the field, which
include data from the Anglo-Australian Observatory (AAO, available
from the web) and data obtained using the VLT-UT1 Test Camera during
the Science Verification period (see ESO home page).  These
comparisons were done by comparing object catalogs extracted using
SExtractor and the zero-points as provided by the original
references. A direct comparison with the AAO data yields in \r-band a
mean difference $<R_{NTT} - R_{AAO}>=$ 0.02~mag and an rms of 0.08~mag
in the range 17.0-22.5. In the \b-band, the mean difference is found
to be 0.12~mag and the rms $\sim$ 0.17~mag, in the magnitude range
16.0-24.0.  Comparing the present data with those in
\r\ band obtained from the VLT-UT1 test camera images, one finds
an rms $\sim$0.13~mag down to \r=25. Unfortunately, there is no
independent zero-point for the \r\ image obtained with the test
camera.

Using the HST-WFPC2 HDF-S observations catalogs were extracted for
each passband at full-resolution, as well as at a resolution comparable
to that of the SUSI2 images (by convolving the WFPC2 image with a
Gaussian with a FWHM of 1 arcsec). These data have been used to further
check the photometric and astrometric calibration of the NTT data and
to evaluate the reliability of the detections. Figure~\ref{fig:wfpc2}
shows the comparison of the magnitudes derived from SExtractor for
sources detected on the convolved WFPC2 image with those measured from
the co-added SUSI2 for each available passband. Also shown are the
errors computed by SExtractor and the scatter, computed in 0.5~mag
bins directly from the comparison of the two data sets. The results
show that while there are differences in zero-points, expected because
of the different passbands of the filters, the scatter is consistent
with the errors estimated by SExtractor.

\begin{figure}[t]
\resizebox{0.40\textwidth}{!}{\includegraphics{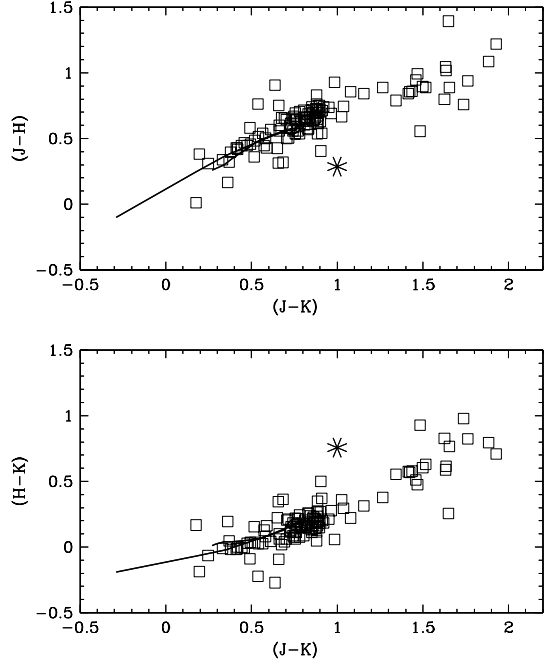}}
\caption{Infrared color-color diagrams for point sources in the HDF2 field
compared with model predictions (solid line). Also shown is the QSO in
the STIS field (star).}
\label{fig:starcolir}
\end{figure}

To test the accuracy of the derived astrometric solution the position
of the objects detected in the $\chi^2$ image, created by combining
the multi-band SUSI2 images, were compared with the HST-WFPC2 images
and the relative shift in positions measured. Relative to the WFPC2
detections, the EIS objects are displaced to the northeast. The
amplitude of the shift is $0.11$~arcsec in right ascension and
$0.4$~arcsec in declination relative to that used by
STScI. Furthermore, based on this comparison the internal accuracy of
the astrometric solution is estimated to be $\lsim$0.4~arcsec.  The
measured shift is consistent with the error of the USNO-A V1.0 catalog
used in the present paper relative to the more accurate astrometric
reference catalog used in the analysis of the HDF-S data.

The results presented in this section demonstrate that the photometric
zero-points. errors in the magnitude measurements and the astrometric
solution obtained in the present paper are reliable.

\begin{figure}[t]
\resizebox{0.38\textwidth}{!}{\includegraphics{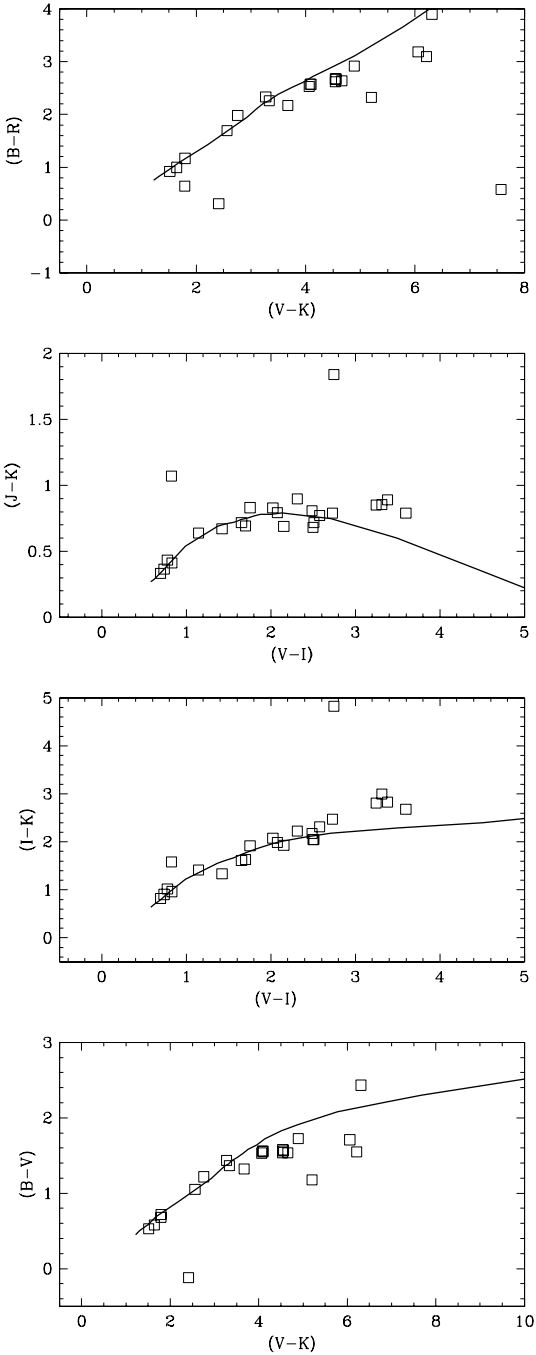}}
\caption{Optical-infrared color-color diagrams for point sources in 
the HDF2 field compared with model predictions (solid line).}
\label{fig:starcoloir}
\end{figure}

\section{Discussion}
\label{results}

For a preliminary evaluation of the overall quality of the photometric
calibration and object detection, photometry and classification, in
this section simple statistics such as galaxy and star number counts,
and color-color diagrams produced from the catalogs described above
are compared with other available data. Since optical data are only
available for HDF2, the following discussion will concentrate on this
region, unless otherwise specified.

\subsection{Point-Sources}

To facilitate comparisons with other data sets, throughout this
section magnitudes are expressed in the Johnson-Cousins system.  Even
though the number of stars in the HDF2 field is relatively small
($\sim80$), it is of interest to compare the star counts with model
predictions, thus providing an indication of the classification
limit. Figure~\ref{fig:ncounts_star} compares the measured star-counts
in the HDF2 field, for each optical passband available, with the
galactic model predictions of the Besancon Observatory group (\eg
Haywood, Robin, \& Creze 1997). Stars are defined to be sources with a
stellarity index $\geq 0.85$ in the \i-band. The observed counts are
in good agreement with the model predictions down to the
classification limit which corresponds $U\lsim 22.5$, $B\lsim 24.0$,
$V\lsim 24.0$, $R\lsim 22.5$, and $I\lsim 22.0$.

The stellar color-color diagrams also offers an important diagnostic
to evaluate the data. From the comparison of the observed stellar
track with other data and/or with model predictions for different
color combinations one is able to detect any systematic errors in the
photometric calibration. Exploration of the multi-dimensional color
space from data in eight passbands may also reveal interesting
population of objects.

From the many possible optical colors combination,
Figure~\ref{fig:starcol} shows three examples involving all the
passbands used.  All the colors have been corrected for reddening and
the $(U-B)$ and $(B-V)$ colors have been corrected for the color term
derived in section~2.2. To minimize contamination by spurious objects
and/or unresolved galaxies, only $5\sigma$ detections, in all
passbands, are included in these diagrams.  For comparison, the
empirical relations compiled by Caldwell \etal (1993) and, whenever
available, a theoretical isochrone taken from Baraffe \etal (1997)
are used because they provide color information in the infrared. The
theoretical model assumes a 10~Gyr, [M/H]$=-1$ population, more
typical for halo stars, and is fine-tuned to model low-mass main
sequence stars. As it can be seen, in all cases the EIS-DEEP data are
in good agreement with the empirical and/or model sequences, with all
the known features of the stellar population of the main components of
the galaxy easily recognizable. For instance, objects with
$(B-V)\sim0.5$ and $(U-B)\sim-0.2$ are low metallicity halo stars near
the turnoff, located at few kpc from the Galactic plane, while the red
population with $(B-V)\sim1.3$ and $(U-B)\sim1.3$ consist of faint
disk M-dwarfs. These two populations are easily seen in the
$(U-B)\times(B-V)$ diagram.

Potentially interesting objects can be isolated in the multicolor
space from their departure from the stellar sequence.  One such an
example is the blue object at $(B-V)\simeq-0.1$ and $(V-I)\simeq0.9$
($V\sim20.1$), lying in a region typical of low-redshift ($z<1.0$)
quasars (\eg Zaggia \etal 1998). This object also shows distinct
optical-infrared colors (see Figure~\ref{fig:starcoloir}).

\begin{figure}[t]
\resizebox{0.42\textwidth}{!}{\includegraphics{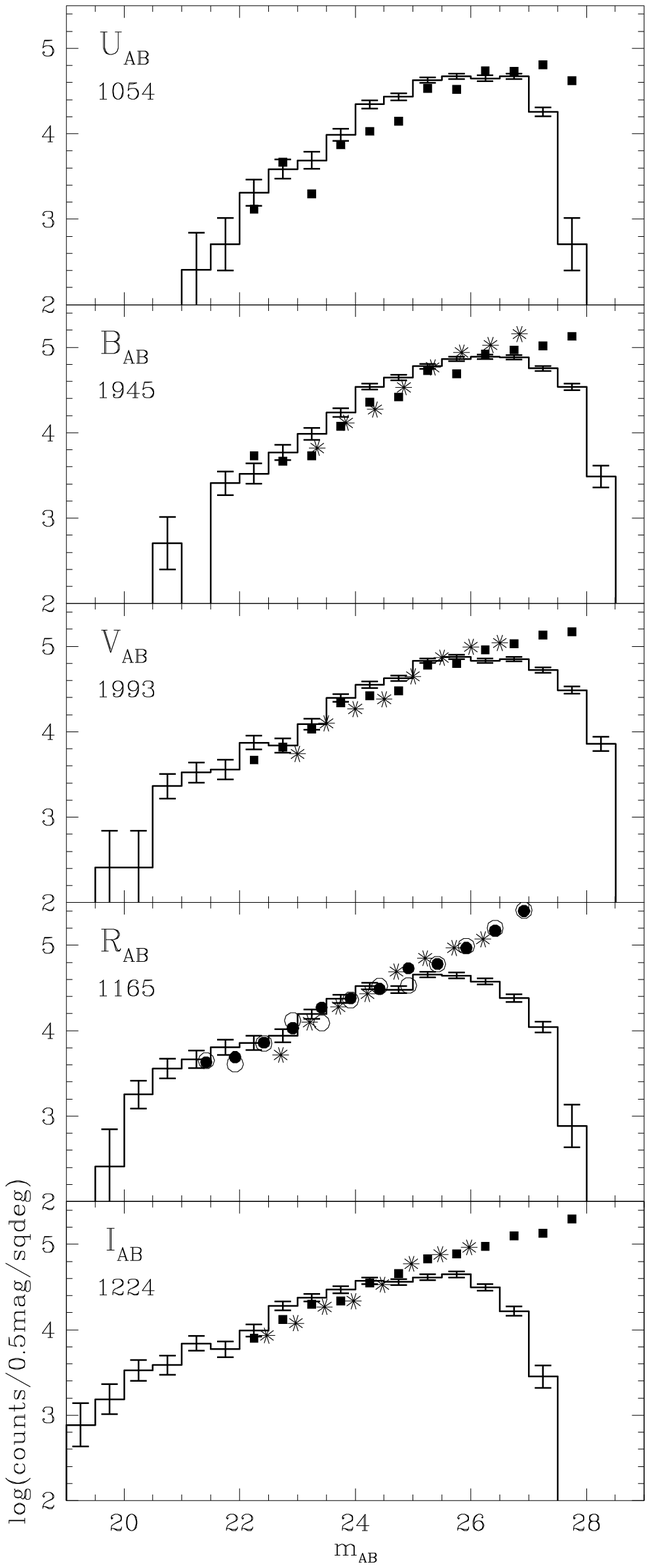}}
\caption{The observed galaxy counts extracted in each of the passbands
used (solid histogram) compared with counts from other authors as
follows: HDF-N counts taken from Williams \etal (1996) (filled
squares); \r-band from Smail \etal (1995) (open circles); \b, \v, and
\i\ galaxy counts from Arnouts \etal (1998) (stars).  The magnitudes
are isophotal and have been transformed to the $AB$ system. The counts
of the other authors have been scaled to a one square degree area.}
\label{ncounts_gal}
\end{figure}

\begin{figure}[t]
\resizebox{0.42\textwidth}{!}{\includegraphics{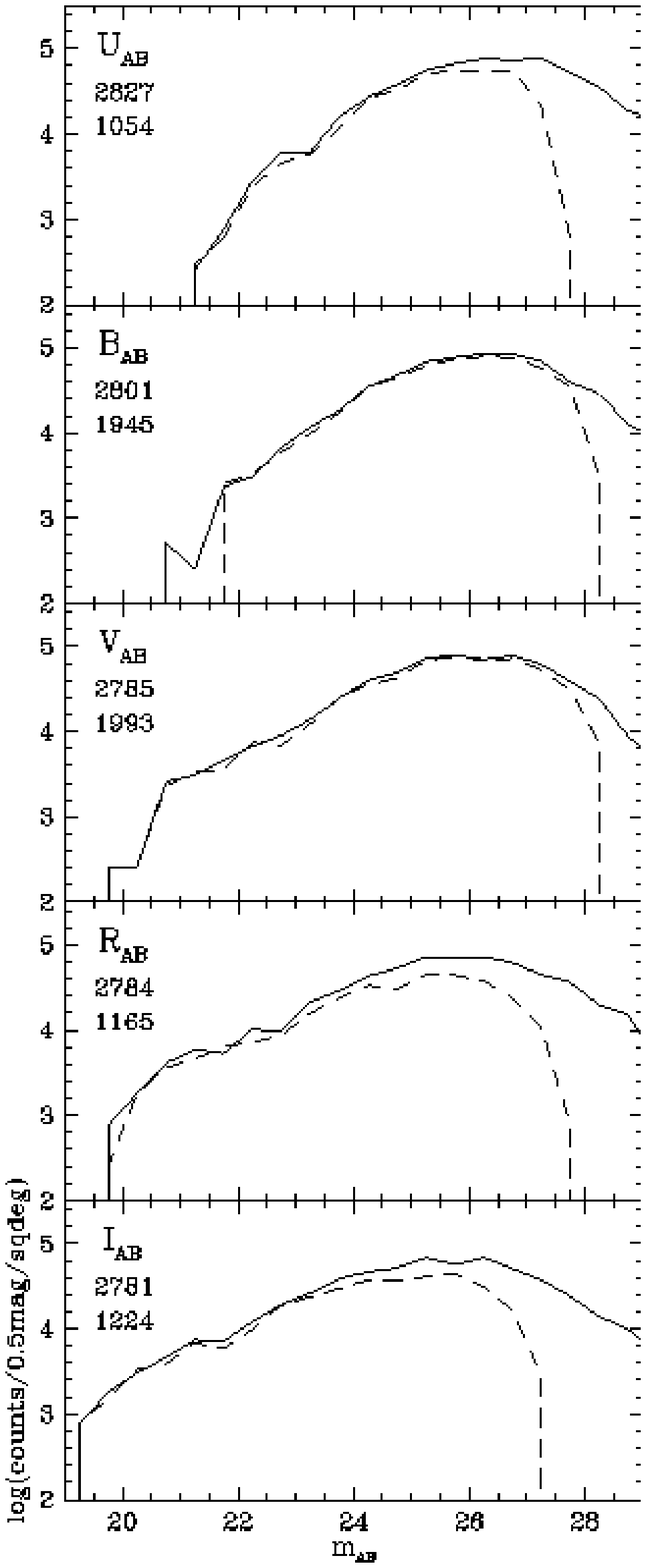}}
\caption{Comparison of the galaxy counts derived in each of the single
passbands (dashed line), with those  derived from the reference
$\chi^2$ image (solid line).}
\label{fig:ncounts_comp}
\end{figure}

Infrared colors for 130 star-like objects with $K\le20$ from HDF1 and
HDF2 fields are shown in Figure~\ref{fig:starcolir}.  Again, the data
are compared to the Baraffe \etal (1997) model. Also shown are the
infrared colors of the QSO ($(J-K)=1.0, (H-K)=0.73$) at $z\sim2.2$
centered in the STIS camera.  Note that the data extends well beyond
the theoretical track which, according to the model, should have a
concentration at $(J-K)\simeq0.8$ and $(J-H)\simeq0.6$. The sparse
population at $(J-K)>1.2$ and $(J-H)\simeq1.0$ are probably unresolved
background galaxies contaminating the stellar sample. In fact, by
setting the limit on the stellarity index to 0.95, most of these
objects disappear from the diagram. Finally, combinations of optical
and infrared colors for 25 star-like objects in the region of overlap
between SUSI2-SOFI are presented in Figure~\ref{fig:starcoloir},
showing again good agreement with the theoretical model.  In summary,
the good agreement between the data and empirical/model stellar in the
optical, infrared and optical-infrared color diagrams are evidences of
the accuracy of the optical and infrared zero-points in the different
passbands used.

\begin{figure}
\resizebox{0.45\textwidth}{!}{\includegraphics{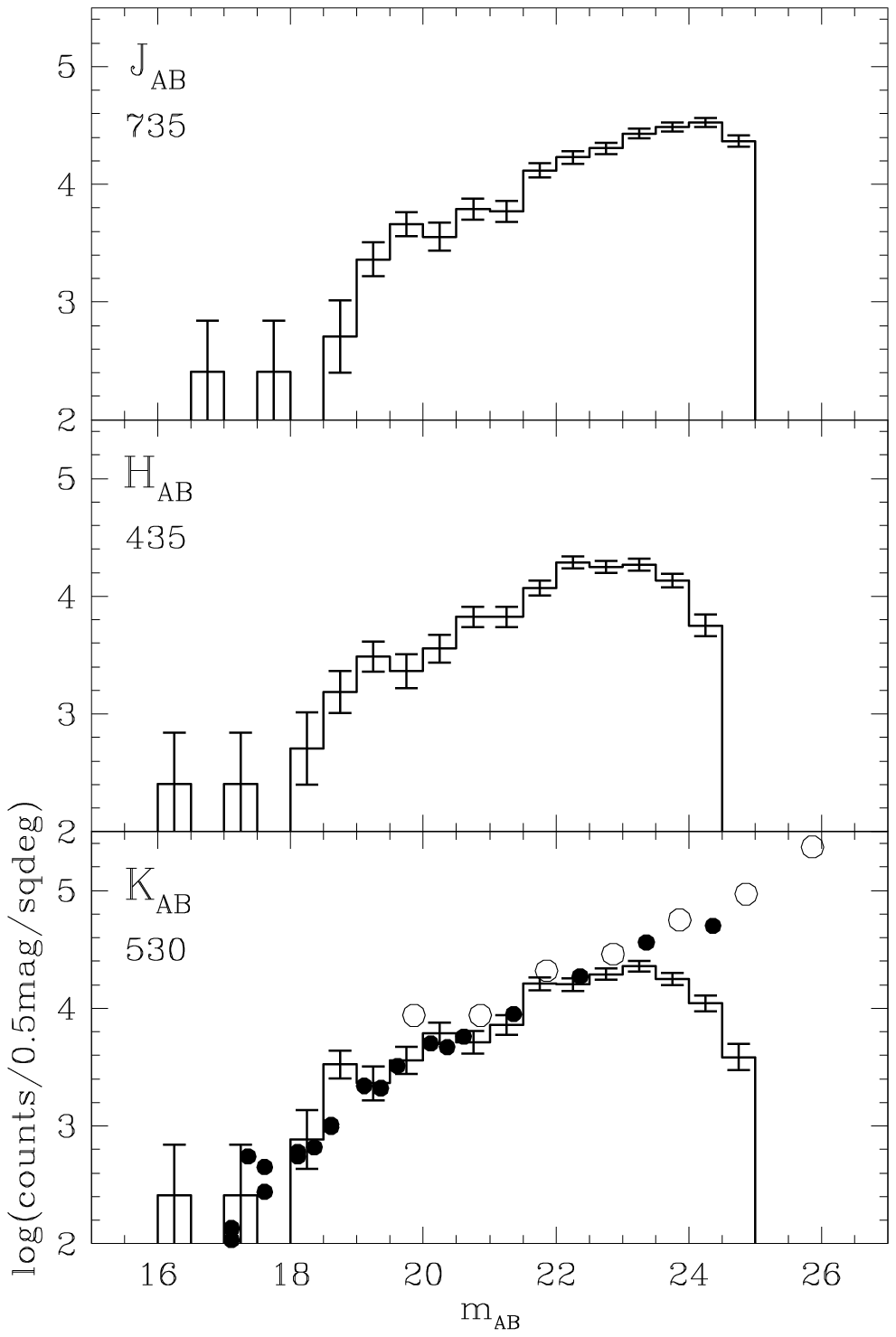}}
\caption{The EIS galaxy counts in the infrared passbands 
compared to those of other authors. As before magnitudes are isophotal 
magnitudes converted to the $AB$ system.}
\label{fig:countsgal_ir}
\end{figure}

\begin{figure}
\resizebox{0.45\textwidth}{!}{\includegraphics{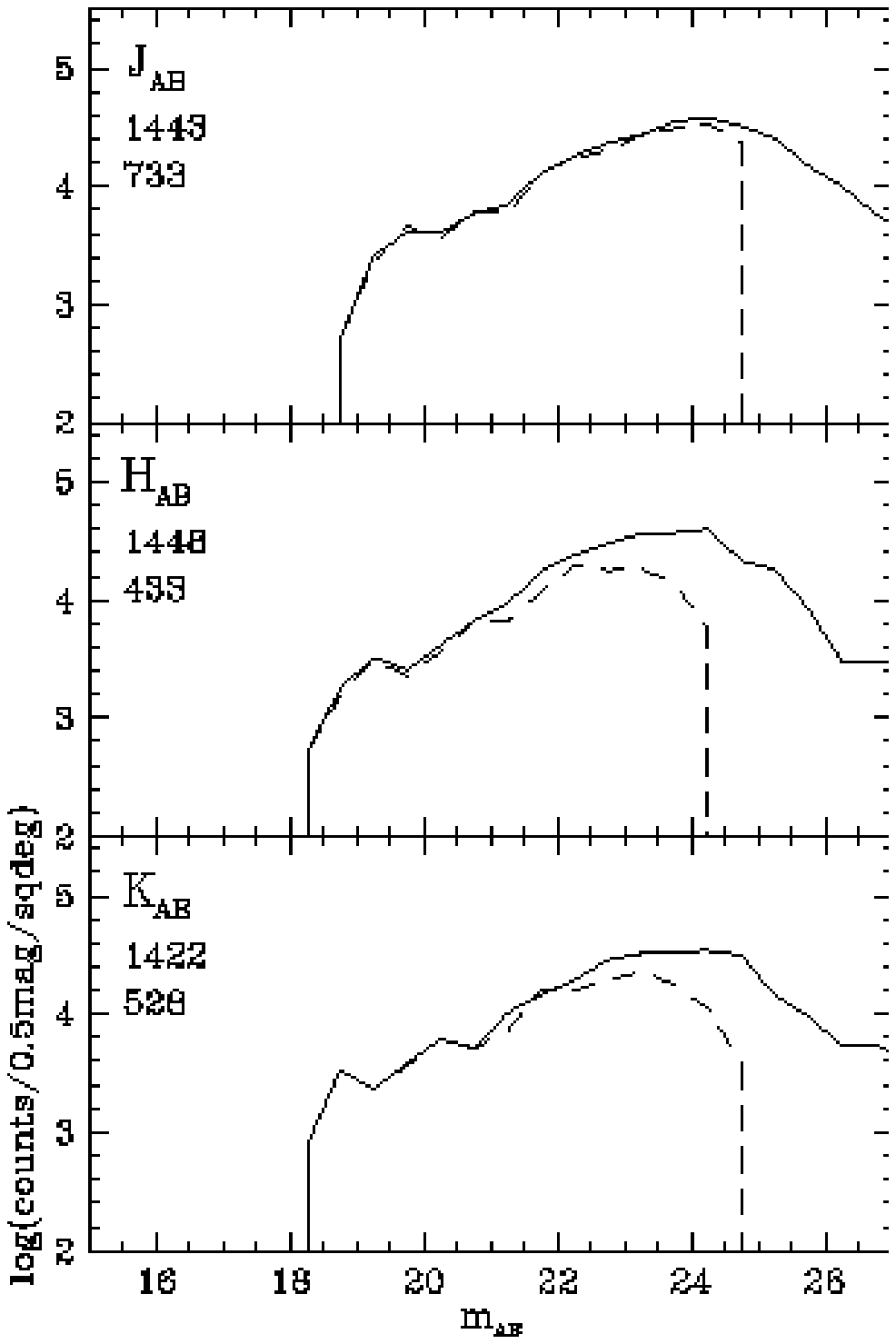}}
\caption{Comparison of the galaxy counts derived in each of the single
passbands (dashed line), with those  derived from the infrared reference
$\chi^2$ image (solid line).}
\label{fig:countsgal_ir_chi2}
\end{figure}

\subsection{Galaxies}

To evaluate the performance and depth of the galaxy catalogs produced
in the different optical passbands, the number counts obtained within
the region defined by the master mask (Section~2.1), with an area of
about $5.3 \times 5.3$ square arcmin, are shown in
Figure~\ref{ncounts_gal}. In the catalogs galaxies are defined to be
objects with a stellarity-index less than 0.85 and SExtractor flags
$<4$.  The counts obtained by other authors from deep imaging of
smaller regions (\eg Smail \etal 1995, Williams \etal 1996, Arnouts
\etal 1998) are given for comparison.  The error-bars in the
histograms are simply the Poisson errors. As can be seen, there is a
remarkable agreement between the EIS-DEEP galaxy-counts and those of
the other authors over a broad range of magnitudes.  The most
significant difference is the observed flattening of the \i-band
counts in the range $I_{AB}$=25-26. This is probably due to the
fringing correction which, as pointed out earlier, may affect the
detections at faint flux levels.  As emphasized earlier the depth in
\u-band is remarkably close to the limit achieved with the HST-WFPC2
camera.  Using the HDF counts as a reference the completeness limit of
the catalogs is roughly $U_{AB}=27$, $B_{AB}=26.7$, $V_{AB}=26$,
$R_{AB}=25.5$ and $I_{AB}=25$. The density of galaxies ranges from 28
to 70 galaxies per square arcmin.

\begin{figure*}
\resizebox{0.95\textwidth}{!}{\includegraphics{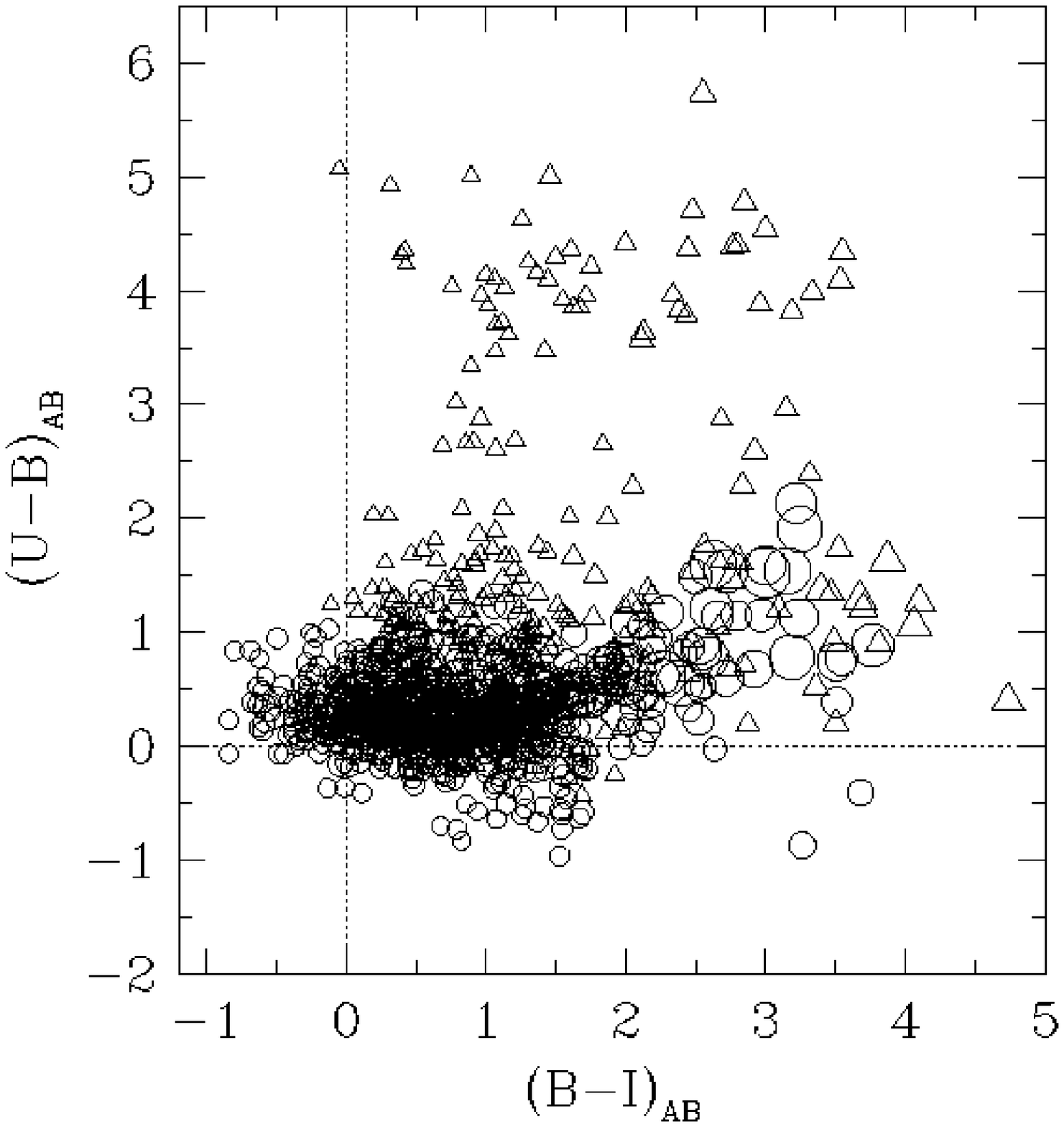}
\includegraphics{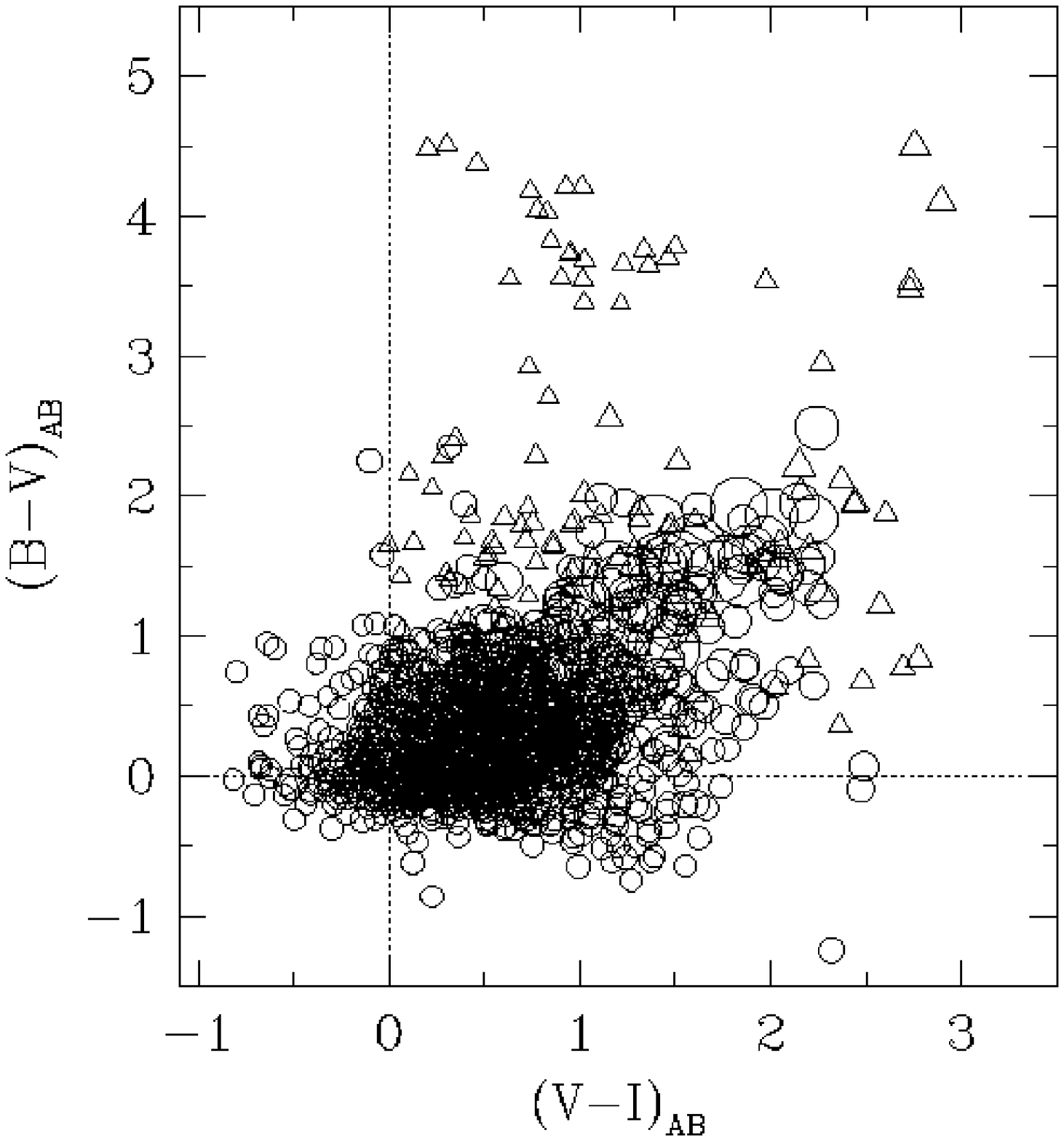}}
\caption{Galaxy $(U-B)\times(B-I)$ and $(B-V)\times(V-I)$ 
color-color diagrams. In the plot the selection boxes proposed by
Madau \etal (1996) to identify galaxies in the redshift ranges $3<z<4$
(left panel) and $z>4$ (right panel) are also shown. The size of
the symbols are inversely proportional to the
\i-band magnitude of the galaxy.}
\label{fig:galcol}
\end{figure*}

In order to evaluate the performance of the $\chi^2$ detection method
used in the production of multi-band catalogs, in
Figure~\ref{fig:ncounts_comp} the distributions of galaxy number counts
obtained from the $\chi^2$ detections are compared with those obtained
in each passband.  As can be seen there is a dramatic increase in the
number of detections, especially in the \u, \r\ and
\i-bands. which in some cases is as large as a  factor of 3. Of
course, these detections are primarily at the faint end.  The
reliability of the detections obtained from the $\chi^2$ image is
impressive. Visual inspection of the detections, using the HST-WFPC2
observations, indicates that the overall reliability of the $\chi^2$
detections is 93\%. For objects with $I<26$, the reliability is better
than 95\%, implying a gain of about 1~mag in the limiting
magnitude. The reliability decreases to 45\% for detections
corresponding to objects for which the measured isophotal \i-band
magnitude in the co-added image is fainter than \i=28.

\begin{table*}[t]
\caption{High-z Galaxy Candidates}
\label{tab:hiz}
\begin{tabular}
{rccc
r@{\extracolsep{2mm}}r@{\hspace{1mm}}
r@{\extracolsep{2mm}}r@{\hspace{1mm}}
r@{\extracolsep{2mm}}r@{\hspace{1mm}}
r@{\extracolsep{2mm}}r@{\hspace{1mm}}
r@{\extracolsep{2mm}}r@{\hspace{1mm}}
r@{\extracolsep{2mm}}r@{\hspace{1mm}}}
\hline\hline\noalign{\smallskip}
\# & 
$\alpha$ &
$\delta$ &
m$_{ap}$ & 
$(U-B)$ & $\epsilon$ &
$(B-V)$ & $\epsilon$ &
$(V-I)$ & $\epsilon$ &
$(B-I)$ & $\epsilon$ &
$(R-I)$ & $\epsilon$ &
$(I-K)$ & $\epsilon$ \\
\noalign{\smallskip}\hline\noalign{\smallskip}
   1	& 22:32:20.42 & $-60$:31:49.9 &  $25.54$ &  $ 0.59$ & $ -1$&$ 1.65$ &  $ 0.57$ &  $ 0.00$ &  $ 0.32$ &  $ 1.65$ &  $ 0.63$ & $
   0.01$ &  $ 0.36$ & -&-\\  
   2	& 22:32:20.83 & $-60$:35:52.5 &  $25.44$ &  $ 3.02$ &  $ -1$ &  $ 0.42$ &  $ 0.25$ &  $ 0.36$ &  $ 0.28$ &  $ 0.79$ &  $ 0.33$ &  $ 0.14$ &  $ 0.31$  & -&-\\  
   3	& 22:32:21.10 & $-60$:34:51.0 &  $25.40$ &  $ 1.55$ &  $ -1$ &  $ 0.54$ &  $ 0.33$ &  $ 0.66$ &  $ 0.29$ &  $ 1.20$ &  $ 0.37$ &  $ 0.13$ &  $ 0.30$  & -&-\\  
   4	&22:32:22.13 & $-60$:35:09.0 &  $25.83$ &  $ 3.96$ &  $-1$ &  $ 0.75$ &  $ 0.34$ &  $ 0.21$ &  $ 0.35$ &  $ 0.96$ &  $ 0.44$ &  $-0.12$ &  $ 0.38$  & -&-\\  
   5	& 22:32:22.40 & $-60$:35:09.3 &  $25.22$ &  $ 1.73$ &  $ 0.93$ &  $ 0.50$ &  $ 0.17$ &  $ 0.05$ &  $ 0.22$ &  $ 0.55$ &  $ 0.25$ &  $-0.23$ &  $ 0.24$  & -&-\\  
   6	& 22:32:22.63 & $-60$:35:29.3 &  $25.95$ &  $-0.37$ &  $-1$ &  $ 3.74$ &  $-1$ &  $ 0.95$ &  $ 0.50$ &  $ 4.69$ &  $-1$ &  $ 0.24$ &  $ 0.51$  & -&-\\  
   7	& 22:32:22.64 & $-60$:33:03.0 &  $26.17$ &  $-2.21$ &  $-1$ &  $ 3.56$ &  $-1$ &  $ 0.90$ &  $ 0.61$ &  $ 4.47$ &  $-1$ &  $-0.13$ &  $ 0.57$  & -&-\\  
   8	& 22:32:22.75 & $-60$:32:59.1 &  $24.91$ &  $ 1.68$ &  $ 1.02$ &  $ 0.47$ &  $ 0.19$ &  $ 0.49$ &  $ 0.19$ &  $ 0.95$ &  $ 0.23$ &  $ 0.19$ &  $ 0.21$  & -&-\\  
   9	& 22:32:22.93 & $-60$:34:29.9 &  $25.38$ &  $ 2.85$ &  $-1$ &  $ 1.56$ &  $ 0.61$ &  $ 0.51$ &  $ 0.28$ &  $ 2.07$ &  $ 0.64$ &  $-0.00$ &  $ 0.29$  & -&-\\  
  10 	& 22:32:26.00 & $-60$:34:51.2 &  $25.82$ &  $ 4.15$ &  $-1$ &  $ 0.89$ &  $ 0.37$ &  $ 0.11$ &  $ 0.38$ &  $ 1.00$ &  $ 0.49$ &  $-0.52$ &  $ 0.38$  & -&-\\  
   11	& 22:32:26.56 & $-60$:35:32.1 &  $25.77$ &  $ 5.01$ &  $-1$ &
   $ 0.42$ &  $ 0.35$ &  $ 0.47$ &  $ 0.38$ &  $ 0.89$ &  $ 0.45$ &  $
   0.32$ &  $ 0.45$ &- &-\\  
   12	& 22:32:27.59 & $-60$:32:58.5 &  $25.61$ &  $ 1.89$ &  $-1$ &  $ 2.41$ &  $ 1.41$ &  $ 0.35$ &  $ 0.34$ &  $ 2.76$ &  $-1$ &  $-0.40$ &  $ 0.34$  & -&-\\  
   13	& 22:32:27.64 & $-60$:33:51.8 &  $26.15$ &  $ 4.04$ &  $-1$ &  $ 0.53$ &  $ 0.42$ &  $ 0.22$ &  $ 0.51$ &  $ 0.76$ &  $ 0.60$ &  $ 0.47$ &  $ 0.68$  & -&-\\  
   14	& 22:32:27.90 & $-60$:34:11.6 &  $25.28$ &  $ 0.58$ &  $ 1.26$ &  $ 1.50$ &  $ 0.49$ &  $ 0.41$ &  $ 0.25$ &  $ 1.92$ &  $ 0.52$ &  $ 0.04$ &  $ 0.27$  & -&-\\  
   15	& 22:32:29.74 & $-60$:35:58.1 &  $25.11$ &  $ 3.86$ &  $-1$ &  $ 1.06$ &  $ 0.35$ &  $ 0.61$ &  $ 0.22$ &  $ 1.67$ &  $ 0.38$ &  $ 0.58$ &  $ 0.28$  & -&-\\  
   16	& 22:32:29.83 & $-60$:35:33.3 &  $26.00$ &  $-0.41$ &  $-1$ &  $ 3.82$ &  $-1$ &  $ 0.85$ &  $ 0.51$ &  $ 4.67$ &  $-1$ &  $ 0.22$ &  $ 0.54$  & -&-\\  
   17	& 22:32:30.68 & $-60$:36:16.2 &  $24.34$ &  $ 2.82$ &  $-1$ &  $ 2.02$ &  $ 0.58$ &  $ 1.02$ &  $ 0.14$ &  $ 3.04$ &  $ 0.58$ &  $ 0.29$ &  $ 0.13$  & -&-\\  
   18	& 22:32:30.95 & $-60$:32:43.8 &  $24.77$ &  $ 0.39$ &  $ 1.20$ &  $ 1.84$ &  $ 0.52$ &  $ 0.61$ &  $ 0.18$ &  $ 2.45$ &  $ 0.53$ &  $ 0.33$ &  $ 0.20$  & -&-\\  
   19	& 22:32:30.99 & $-60$:36:16.7 &  $25.28$ &  $ 2.56$ &  $-1$ &  $ 1.68$ &  $ 0.74$ &  $ 0.72$ &  $ 0.26$ &  $ 2.40$ &  $ 0.76$ &  $ 0.13$ &  $ 0.27$  & -&-\\  
   20	& 22:32:31.00 & $-60$:35:32.1 &  $25.52$ &  $-0.38$ &  $-1$ &  $ 4.21$ &  $-1$ &  $ 0.93$ &  $ 0.35$ &  $ 5.14$ &  $-1$ &  $ 0.33$ &  $ 0.37$  & -&-\\  
   21	& 22:32:31.07 & $-60$:33:29.7 &  $25.49$ &  $ 2.68$ &  $-1$ &  $ 0.89$ &  $ 0.27$ &  $ 0.02$ &  $ 0.29$ &  $ 0.91$ &  $ 0.36$ &  $-0.35$ &  $ 0.30$  & -&-\\  
   22	& 22:32:31.28 & $-60$:33:25.3 &  $24.94$ &  $ 4.36$ &  $-1$ &  $ 1.34$ &  $ 0.29$ &  $ 0.26$ &  $ 0.19$ &  $ 1.61$ &  $ 0.33$ &  $ 0.07$ &  $ 0.21$  & -&-\\  
   23	& 22:32:31.87 & $-60$:35:15.9 &  $24.23$ &  $-0.39$ &  $ 1.34$ &  $ 2.56$ &  $ 0.94$ &  $ 1.16$ &  $ 0.14$ &  $ 3.71$ &  $ 0.94$ &  $ 0.61$ &  $ 0.14$  & -&-\\  
   24	& 22:32:32.56 & $-60$:34:14.8 &  $24.96$ &  $ 4.22$ &  $-1$ &  $ 0.76$ &  $ 0.35$ &  $ 0.99$ &  $ 0.23$ &  $ 1.75$ &  $ 0.36$ &  $ 0.13$ &  $ 0.21$  & -&-\\  
   25	& 22:32:33.30 & $-60$:31:37.2 &  $25.24$ &  $ 0.24$ &  $-1$ &  $ 4.48$ &  $ -1$ &  $ 0.21$ &  $ 0.24$ &  $ 4.68$ &  $ -1$ &  $ 0.34$ &  $ 0.30$  & -&-\\  
   26	& 22:32:34.08 & $-60$:36:13.3 &  $25.35$ &  $ 4.10$ &  $-1$ &  $ 0.63$ &  $ 0.38$ &  $ 0.81$ &  $ 0.29$ &  $ 1.44$ &  $ 0.41$ &  $-0.15$ &  $ 0.27$  & -&-\\  
   27	& 22:32:34.98 & $-60$:33:36.5 &  $25.17$ &  $ 2.65$ &  $ -1$ &  $ 0.89$ &  $ 0.45$ &  $ 0.94$ &  $ 0.27$ &  $ 1.83$ &  $ 0.46$ &  $ 0.26$ &  $ 0.27$  & -&-\\  
   28	& 22:32:35.92 & $-60$:34:54.4 &  $25.42$ &  $ 4.25$ &  $-1$ &  $ 0.87$ &  $ 0.35$ &  $ 0.44$ &  $ 0.29$ &  $ 1.30$ &  $ 0.40$ &  $ 0.62$ &  $ 0.39$  & -&-\\  
   29	& 22:32:35.96 & $-60$:32:11.8 &  $25.09$ &  $ 4.30$ &  $-1$ &  $ 0.84$ &  $ 0.33$ &  $ 0.65$ &  $ 0.23$ &  $ 1.50$ &  $ 0.36$ &  $ 0.30$ &  $ 0.26$  & -&-\\  
   30	& 22:32:36.28 & $-60$:36:16.1 &  $24.50$ &  $ 4.42$ &  $-1$ &  $ 0.98$ &  $ 0.29$ &  $ 1.01$ &  $ 0.16$ &  $ 2.00$ &  $ 0.29$ &  $ 0.49$ &  $ 0.17$  & -&-\\  
   31	& 22:32:38.55 & $-60$:36:29.0 &  $25.33$ &  $ 4.16$ &  $-1$ &  $ 0.27$ &  $ 0.45$ &  $ 1.09$ &  $ 0.37$ &  $ 1.36$ &  $ 0.46$ &  $ 0.95$ &  $ 0.52$  & -&-\\  
   32	& 22:32:38.61 & $-60$:34:13.4 &  $25.27$ &  $ 1.99$ &  $-1$ &  $ 2.29$ &  $ -1$ &  $ 0.77$ &  $ 0.30$ &  $ 3.06$ &  $ -1$ &  $-0.00$ &  $ 0.30$  & -&-\\  
   33	& 22:32:40.05 & $-60$:36:21.1 &  $24.10$ &  $ 1.67$ &  $ 1.04$ &  $ 1.13$ &  $ 0.18$ &  $ 0.50$ &  $ 0.12$ &  $ 1.62$ &  $ 0.20$ &  $ 0.24$ &  $ 0.13$  & -&-\\  
   34	& 22:32:40.29 & $-60$:35:25.4 &  $25.86$ &  $ 2.36$ &  $-1$ &  $ 1.67$ &  $ 0.81$ &  $ 0.13$ &  $ 0.47$ &  $ 1.80$ &  $ 0.91$ &  $ 0.45$ &  $ 0.63$  & -&-\\  
   35	& 22:32:41.03 & $-60$:34:04.6 &  $25.79$ &  $-0.48$ &  $-1$ &  $ 3.69$ &  $-1$ &  $ 1.02$ &  $ 0.53$ &  $ 4.72$ &  $-1$ &  $ 0.12$ &  $ 0.51$  & -&-\\  
   36 & 22:32:42.19 & $-60$:34:46.1 &  $23.98$ &  $ 2.45$ &  $-1$ &  $ 1.13$ &  $ -1$ &  $ 4.84$ &  $ -1$ &  $ 5.97$ &  $ 
-1$ &  $ 1.80$ &  $ 0.30$ &  $ 1.10$ &  $ 0.20$ \\  
   37	& 22:32:42.42 & $-60$:35:49.3 &  $25.59$ &  $ 2.26$ &  $-1$ &  $ 1.64$ &  $ 0.94$ &  $ 0.56$ &  $ 0.38$ &  $ 2.20$ &  $ 0.98$ &  $ 0.45$ &  $ 0.47$ &- &-  \\  
   38	& 22:32:44.51 & $-60$:33:12.6 &  $25.51$ &  $ 1.69$ &  $ 1.24$ &  $ 0.46$ &  $ 0.22$ &  $ 0.00$ &  $ 0.30$ &  $ 0.46$ &  $ 0.33$ &  $-0.04$ &  $ 0.33$  & -&-\\  
   39   & 22:32:45.03 & $-60$:34:56.1 &  $25.70$ &  $ 1.81$&  $ -1$ &  $-0.04$ &  $ 0.43$ &  $ 0.68$ &  $ 0.51$ &  $ 0.64$ &
   $ 0.50$ &  $ 0.04$ &  $ 0.49$ &$-0.05$ &  $ 1.52$ \\  
   40	& 22:32:45.05 & $-60$:31:27.6 &  $25.65$ &  $-0.42$ &  $-1$ &  $ 3.66$ &  $-1$ &  $ 1.23$ &  $ 0.45$ &  $ 4.89$ &  $-1$ &  $-0.17$ &  $ 0.36$  & -&-\\  
   41   & 22:32:45.24 & $-60$:35:05.1 &  $25.11$ &  $ 4.18$ &  $-1$ &  $ 0.40$ &  $ 0.49$ &  $ 1.20$ &  $ 0.38$ &  $ 1.60$ &  $ 0.47$ &  $ 0.26$ &  $ 0.32$ & $ 0.77$ &  $ 0.52$ \\  
  42	 & 22:32:46.72 & $-60$:32:10.7 &  $25.93$ &  $ 2.64$ &  $ -1$ &  $ 0.46$ &  $ 0.37$ &  $ 0.23$ &  $ 0.44$ &  $ 0.69$ &  $ 0.51$ &  $-0.21$ &  $ 0.46$  & -&-\\  
  43   & 22:32:46.99 & $-60$:31:46.5 &  $24.64$ &  $ 2.17$ &  $ 1.16$ &  $ 0.64$ &  $ 0.20$ &  $ 0.27$ &  $ 0.20$ &  $ 0.91$ &  $ 0.24$ &  $ 0.11$ &  $ 0.21$ & $ 0.70$ &  $ 0.38$ \\  
  44	 & 22:32:47.02 & $-60$:36:02.5 &  $25.67$ &  $ 2.14$ &  $-1$ &  $ 1.79$ &  $ 1.17$ &  $ 0.69$ &  $ 0.37$ &  $ 2.48$ &  $ 1.19$ &  $-0.08$ &  $ 0.37$  & -&-\\  
  45   & 22:32:47.49 & $-60$:35:09.2 &  $25.67$ &  $-3.14$ &  $-1$ &  $ 5.02$ &  $-1$ &  $ 1.11$ &  $ 0.59$ &  $ 6.12$ &  $-1$ &  $ 0.66$ &  $ 0.61$ &  $-2.87$ &  $-1$ \\  
  46   & 22:32:48.45 & $-60$:32:37.3 &  $25.02$ &  $ 6.07$ &  $-1$ &
   $ 0.76$ &  $ 0.37$ &  $ 0.66$ &  $ 0.29$ &  $ 1.42$ &  $ 0.40$ &  $
   0.06$ &  $ 0.28$ &  $ 1.49$ &  $ 0.35$ \\ 
   47 & 22:32:49.33 & $-60$:32:25.3 &  $23.44$ &  $ 1.52$ &  $ 0.30$ &  $ 0.46$ &  $ 0.09$ &  $ 0.20$ &  $ 0.09$ &  $ 0.65$ &  $ 0
.10$ &  $ 0.02$ &  $ 0.09$ &  $ 0.40$ &  $ 0.20$ \\  
  48   & 22:32:49.97 & $-60$:34:47.1 &  $25.80$ &  $-0.10$&  $-1$ &  $ 5.61$ &  $-1$ &  $ 0.38$ &  $ 0.52$ &  $ 5.99
$ &  $-1$ &  $ 0.00$ &  $ 0.53$ &  $ 0.49$ &  $ 1.10$ \\  
   49   & 22:32:50.31 & $-60$:31:27.0 &  $25.64$ &  $-0.10$ &  $-1$ &  $ 5.24$ &  $-1$ &  $ 0.91$ &  $ 0.53$ &  $ 6.16$ &  $-1$ &  $ 0.58$ &  $ 0.57$ &  $ 1.43$ &  $ 0.57$ \\ 
  50	 & 22:32:51.05 & $-60$:35:09.1 &  $25.40$ &  $ 3.86$ &  $-1$ &  $ 1.11$ &  $ 0.45$ &  $ 0.52$ &  $ 0.29$ &  $ 1.62$ &  $ 0.49$ &  $ 0.16$ &  $ 0.32$  & -&-\\  
   51    & 22:32:52.07 & $-60$:33:42.1 &  $24.98$ &  $ 6.13$ &  $-1$ &  $ 0.73$ &  $ 0.33$ &  $ 0.56$ &  $ 0.28$ &  $ 1.29$ &  $ 0.36$ &  $ 0.11$ &  $ 0.27$ & $ 1.00$ &  $ 0.42$ \\  
\hline\hline
\end{tabular}

\end{table*}
\begin{table*}[h]
\addtocounter{table}{-1}
\caption{High-z Galaxy Candidates. Continued.}
\begin{tabular}
{rccc
r@{\extracolsep{2mm}}r@{\hspace{1mm}}
r@{\extracolsep{2mm}}r@{\hspace{1mm}}
r@{\extracolsep{2mm}}r@{\hspace{1mm}}
r@{\extracolsep{2mm}}r@{\hspace{1mm}}
r@{\extracolsep{2mm}}r@{\hspace{1mm}}
r@{\extracolsep{2mm}}r@{\hspace{1mm}}}
\hline\hline\noalign{\smallskip}
\# & 
$\alpha$ &
$\delta$ &
m$_{ap}$ & 
$(U-B)$ & $\epsilon$ &
$(B-V)$ & $\epsilon$ &
$(V-I)$ & $\epsilon$ &
$(B-I)$ & $\epsilon$ &
$(R-I)$ & $\epsilon$ &
$(I-K)$ & $\epsilon$ \\
\noalign{\smallskip}\hline\noalign{\smallskip}
  52	 & 22:32:53.13 & $-60$:32:06.2 &  $24.36$ &  $ 2.01$ &  $ -1$ &  $ 1.04$ &  $ 0.23$ &  $ 0.82$ &  $ 0.13$ &  $ 1.87$ &  $ 0.24$ &  $ 0.30$ &  $ 0.13$  & -&-\\  
   53    & 22:32:53.46 & $-60$:35:24.5 &  $24.46$ &  $ 4.85$ &  $-1$ &  $ 1.57$ &  $ 0.45$ &  $ 0.79$ &  $ 0.20$ &  $ 2.36$ &  $ 0.45$ &  $-0.01$ &  $ 0.18$ &  $ 1.23$ &  $ 0.26$ \\  
   54    & 22:32:53.48 & $-60$:33:11.7 &  $24.83$ &  $ 2.28$ &  $-1$ &  $ 0.40$ &  $ 0.27$ &  $ 0.69$ &  $ 0.25$ &  $ 1.08$ &  $ 0.30$ &  $ 0.47$ &  $ 0.28$ & $ 1.22$ &  $ 0.34$ \\  
  55	 & 22:32:53.62 & $-60$:31:58.8 &  $25.87$ &  $ 2.04$ &  $-1$ &  $ 1.53$ &  $ 1.26$ &  $ 0.77$ &  $ 0.46$ &  $ 2.30$ &  $ 1.28$ &  $-0.40$ &  $ 0.42$  & -&-\\  
   56    & 22:32:53.87 & $-60$:35:41.1 &  $25.93$ &  $ 5.76$ &  $-1$ &  $ 0.15$ &  $ 0.50$ &  $ 0.54$ &  $ 0.60$ &  $ 0.70$ &  $ 0.62$ &  $ 0.54$ &  $ 0.72$ &  $-2.62$ &  $-1$ \\  
  57	 & 22:32:54.56 & $-60$:33:23.9 &  $26.02$ &  $ 2.09$ &  $ -1$ &  $ 0.82$ &  $ 0.41$ &  $ 0.00$ &  $ 0.46$ &  $ 0.83$ &  $ 0.57$ &  $ 0.04$ &  $ 0.53$  & -&-\\  
   58    & 22:32:54.68 & $-60$:34:30.6 &  $24.69$ &  $ 4.17$ &  $ -1$ &  $ 0.51$ &  $ 0.22$ &  $ 0.45$ &  $ 0.22$ &  $ 0.96$ &  $ 0.26$ &  $-0.02$ &  $ 0.21$ &  $ 0.92$ &  $ 0.35$ \\  
  59	 & 22:32:55.34 & $-60$:35:42.9 &  $25.54$ &  $-2.50$ &  $-1$ &  $ 3.76$ &  $-1$ &  $ 1.34$ &  $ 0.42$ &  $ 5.10$ &  $-1$ &  $ 0.72$ &  $ 0.45$  & -&-\\  
  60 & 22:32:55.41 & $-60$:33:54.5 &  $23.85$ &  $ 0.22$ &  $ -1$ &  $ 1.21$ &  $ 1.27$ &  $ 2.99$ &  $ 0.46$ &  $ 4.20$ &  $ 1
.19$ &  $ 1.57$ &  $ 0.23$ &  $ 0.94$ &  $ 0.20$ \\  
  61	 & 22:32:55.81 & $-60$:32:51.6 &  $26.18$ &  $ 4.23$ &
   $-1$ &  $ 0.80$ &  $ 0.34$ &  $-0.38$ &  $ 0.52$ &  $ 0.43$ &
   $ 0.59$ &  $-0.14$ &  $ 0.58$  & -&-\\  
   62    & 22:32:56.41 & $-60$:31:43.8 &  $25.55$ &  $ 6.49$ &  $-1$ &  $ 0.77$ &  $ 0.25$ &  $-0.42$ &  $ 0.38$ &  $ 0.35$ &  $ 0.42$ &  $-0.83$ &  $ 0.38$ &  $ 1.81$ &  $ 0.47$ \\  
  63	 & 22:32:58.07 & $-60$:34:55.8 &  $25.90$ &  $-0.18$ &  $-1$ &  $ 3.38$ &  $-1$ &  $ 1.02$ &  $ 0.51$ &  $ 4.40$ &  $-1$ &  $ 0.76$ &  $ 0.64$  & -&-\\  
  64    & 22:32:58.46 & $-60$:34:36.4 &  $25.48$ &  $-0.10$ &  $-1$ &  $ 5.36$ &  $-1$ &  $ 0.96$ &  $ 0.47$ &  $ 6.31$ &  $-1$ &  $ 0.36$ &  $ 0.45$ &  $ 0.61$ &  $ 0.78$ \\  
  65    & 22:32:58.55 & $-60$:31:58.4 &  $25.88$ &  $ 3.67$ &  $-1$ &  $ 1.54$ &  $ 1.21$ &  $ 0.60$ &  $ 0.58$ &  $ 2.14$ &  $ 1.25$ &  $ 0.50$ &  $ 0.67$ &  $-2.67$ &  $-1$ \\  
  66    & 22:32:59.15 & $-60$:35:07.6 &  $25.50$ &  $-0.10$
   &  $-1$ &  $ 4.89$ &  $-1$ &  $ 1.41$ &  $ 0.59$ &  $ 6.30$
   &  $-1$ &  $ 0.88$ &  $ 0.59$ &  $ 1.50$ &  $ 0.50$ \\  
   67   & 22:32:59.21 & $-60$:33:57.2 &  $25.11$ &  $ 4.44$ &  $-1$ &  $ 0.07$ &  $ 0.38$ &  $ 1.06$ &  $ 0.36$ &  $ 1.13$ &  $ 0.37$ &  $ 0.94$ &  $ 0.43$ & $ 2.05$ &  $ 0.31$ \\  
  68	 & 22:33:02.49 & $-60$:31:54.8 &  $25.11$ &  $ 1.89$ &  $ -1$ &  $-0.25$ &  $ 0.37$ &  $ 1.32$ &  $ 0.34$ &  $ 1.07$ &  $ 0.34$ &  $ 1.02$ &  $ 0.43$  & -&-\\  
  69	 & 22:33:02.79 & $-60$:33:32.4 &  $25.90$ &  $-1.20$ &  $
-1$ &  $ 2.28$ &  $ -1$ &  $ 0.28$ &  $ 0.51$ &  $ 2.56$ &  $
-1$ &  $-0.16$ &  $ 0.54$  & -&-\\    
\hline\hline
\end{tabular}

\end{table*}

Similarly, in Figure~\ref{fig:countsgal_ir} the counts in the infrared
passbands are shown combining the two fields observed, based both on
detections in single bands and in the $\chi^2$ image built from the
combination of the $JHKs$ images. As in the case of the optical
catalogs, the multi-color catalogs extracted from the $\chi^2$ image
extend to much fainter magnitudes and have an excellent reliability.

Since one of the primary goals of the survey has been to identify
candidate galaxies at high-z for follow-up spectroscopic observations
with the VLT, the color information available in optical, infrared and
optical-infrared has been used to identify preliminary candidates.
Extensive work has been done to tune the color-selection criteria for
the HST filters (\eg Madau \etal 1996) and to identify regions in the
color-color diagrams populated by Lyman-break galaxies in different
redshift ranges.  However, the differences in the passbands between
the SUSI2 and WFPC2 filters prevent using the same color criteria. It
is important to emphasize that, in general, any U-dropouts in the
filters used by SUSI2 would lead to the identification of galaxies at
larger redshifts because of the shift to red of the SUSI2  passband, and
consequently to a smaller density of objects than that inferred by Madau
\etal (1996). However, given the great interest in finding likely
candidates at high-z a simple approach has been adopted here for a
first cut analysis.  This was done by considering the tracks most
likely to trace the evolutionary sequence of galaxies of different
types in color-color diagrams appropriate for the SUSI2 filters
(Arnouts 1998, see also Fontana \etal 1998).  Based on these results
conservative regions in $(U-B) \times (B-I)$ and $(B-V) \times (V-I)$
diagrams, shown in Figure~\ref{fig:galcol}, were defined.  The
criteria adopted were $(U-B)_{AB}\gsim 1.5$ and $(B-I)_{AB}\lsim 2$ in
the $(U-B) \times (B-I)$ diagram and $(B-V)_{AB}>1.5$ and $(B-V)_{AB}
> 2\times(B-I)_{AB} - 0.14$. Based on the model predictions these
regions should be populated by $z>3$ galaxies.  A more precise
selection will require a more detailed analysis using the color
information to assign photometric redshifts. This will certainly be
pursued by several groups using the public data.

The galaxies shown in Figure~\ref{fig:galcol} are detections obtained
from the optical $\chi^2$ image, thus allowing for the identification
of objects that may be undetected in one or more passbands.  The
objects shown in the color-color diagrams follow several
constraints. First, they have to be $\ge 2\sigma$ detections at least
in $B$ and $I$ for the diagram shown in the left panel and $V$ and $I$
for that in the right panel.  If the object is less than a $2\sigma$
detection in the bluest passband, it is represented as a triangle,
otherwise as a circle.  For blue dropouts the magnitude measured by
SExtractor was used regardless of its error. If the magnitude is not
measurable (m=99.9), the 2$\sigma$ limiting magnitude is assigned to
the object.

\begin{figure*}
\resizebox{0.9\textwidth}{!}{\includegraphics{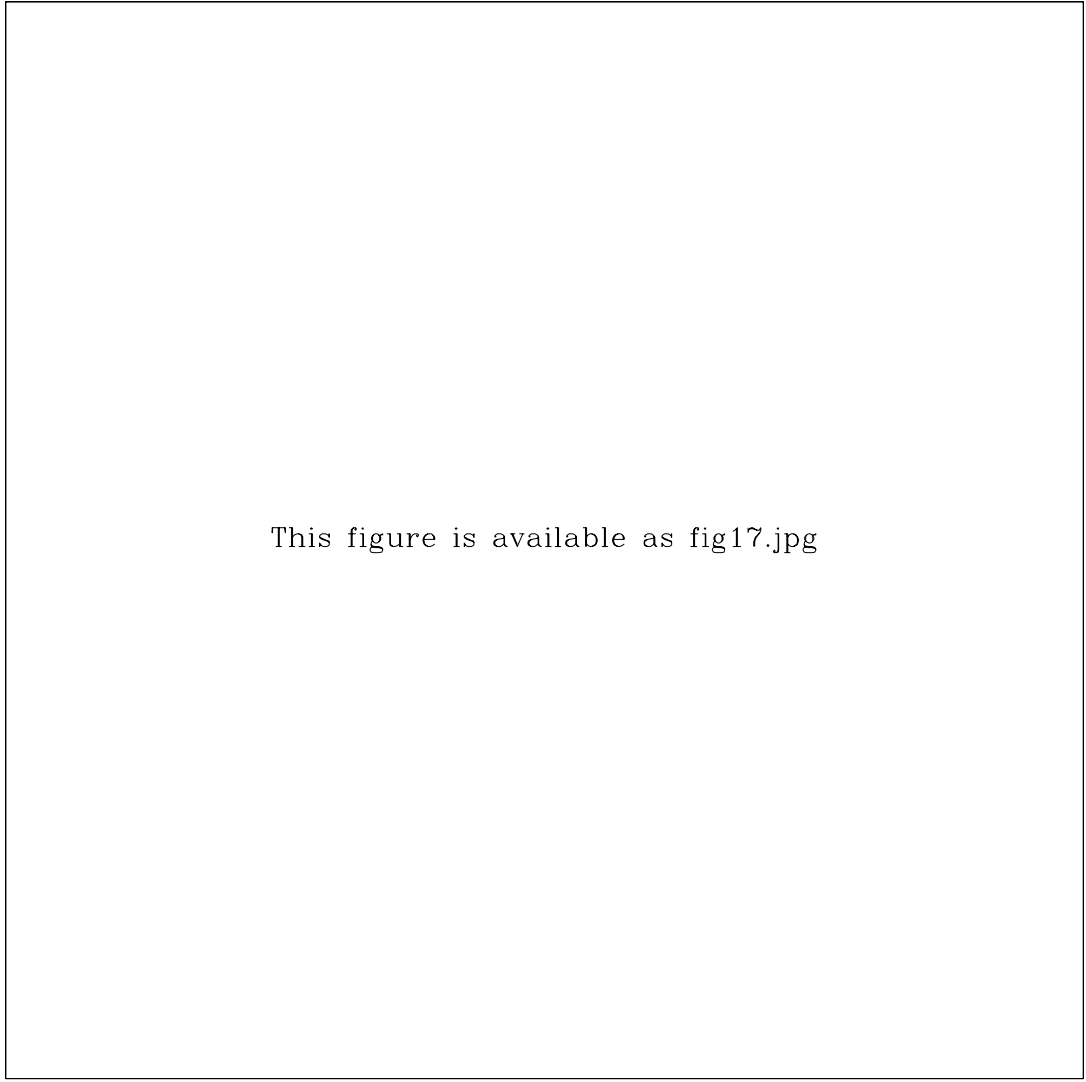}}
\caption{Examples of objects selected as likely high-redshift
galaxies. Each row shows postage stamps extracted from the images in
different passbands showing $UBVIR$ from left to right. The selected
objecta are at the center of each postage stamp.}
\label{fig:drop}
\end{figure*}

Adopting this procedure galaxies the HDF2 field covering $\sim$ 28.1
square arcmin were selected. A total of about 120 candidates
satisfied at least one of the criteria mentioned above. Given the
large uncertainties in the adopted procedure no attempt has been made
to assign these objects to any redshift range.  All candidates were
visually inspected and about half of them were discarded for the
following reasons: 1) close to edge of the frame; 2) near relatively
bright stars; 3) close to masked out regions; and 4) lying along
spikes from the bright stars in and near the field. The remaining
objects, likely to be promising candidates, are listed in
Table~\ref{tab:hiz}.  The table provides the following information: in
columns (1) and (2) right ascension and declination (J2000.0); in
column (3) the $I_{AB}$ magnitude. The remaining columns give the
colors and their respective errors. Note that there are 20 galaxies
within the region of overlap between SUSI2 and SOFI (12.8 square
arcmin), and for them the optical-infrared $(I-K)$ color is also
provided. Whenever, an object is undetected in a given passband or the
$S/N<1$ the error in the color involving this filter is set to -1. In
Figure~\ref{fig:drop} some examples of likely high-redshift galaxies
are shown to illustrate the type of candidates that have been
selected. Finally, note that inspection of the $(V-I)\times (I-K)$
diagram allowed to isolate one possible V-dropout candidate (number 36
in the table), shown in the last row of Figure~\ref{fig:drop}. Even
though the selection criteria adopted are admittedly crude, inspection
of the selected objects indicates that they are by and large promising.

\section{Summary}
\label{sum}

This paper presents the results of deep optical ($\sim$ 30 square
arcmin) and infrared ($\sim$ 40 square arcmin) imaging of two regions
which include the HST-WFPC2 and STIS fields. The former has been
covered in eight passbands from the optical to the infrared, while the
latter in two optical and three infrared passbands. The observations
were carried out as part of the EIS public survey. In addition
single-passband catalogs have been prepared as well as multi-color
optical, infrared and optical-infrared catalogs. The latter were
produced using the $\chi^2$ technique. Preliminary results have shown
that the method is rather promising leading to robust detections of
very faint galaxies.  The color information has been used to find
possible high-z galaxies for follow-up observations with the VLT.  All
the data presented here, in the form of images and catalogs, are being
made public world-wide and can be requested at
"http://www.eso.org/eis''. It is expected that these EIS-DEEP data,
covering a larger area and with deep infrared coverage, will further
contribute to future studies of the HDF-S region.

\begin{acknowledgements}

We thank all the people directly or indirectly involved in the ESO
Imaging Survey effort. In particular, all the members of the EIS
Working Group (S. Charlot, G. Chincarini, S. Cristiani, J. Krautter,
K. Kuijken, K. Meisenheimer, D. Mera, Y. Mellier, M. Ramella, H.
Rottering and P. Schneider) for the innumerable suggestions and
criticisms, the ESO OPC, the NTT team, in particular the night
assistants, the ESO Archive Groups and ECF. We would also like to
thank S. Arnouts, J. Caldwell, N. Devillard, A. Fontana and R. Fosbury
for their help and assistance.  Special thanks to Riccardo Giacconi
for making this effort possible.

\end{acknowledgements}

\end{document}